\documentclass[12pt]{article}
\usepackage{amsmath,amsthm}
\usepackage{bm,amssymb,mathtools}
\usepackage[ruled,vlined]{algorithm2e}
\usepackage{graphicx,psfrag,epsf}
\usepackage{enumerate}
\usepackage[round,authoryear]{natbib}
\usepackage{url} 
\usepackage[colorlinks,citecolor=blue,urlcolor=blue,linkcolor = blue,backref=page]{hyperref}
\usepackage{float}
\usepackage{siunitx}
\usepackage{subcaption}
\usepackage{multirow}
\usepackage{booktabs}
\usepackage{tikz}

\title{Estimating Causal Attribution of Anthropogenic Forcing on High-Temperature Extremes Using \\a Latent Gaussian Spatial Model}
\author{Ritik Roshan Giri and Arnab Hazra \vspace{2mm} \\
Department of Mathematics and Statistics, \\ Indian Institute of Technology Kanpur, Kanpur, India 208016 \vspace{2mm}}

\date{}
\begin{document}

\maketitle

\begin{abstract}
\noindent 
Climate change has become a significant global concern due to its capacity to cause substantial disruption to daily life by increasing the frequency and intensity of extreme weather events. Given the rising trend of human interventions in the climate system over recent decades, this study aims to quantify the relative contribution of anthropogenic forcing to the increasing likelihood of climate extremes, with a particular emphasis on high-temperature extremes. Our analysis focuses on annual temperature maxima from the IPSL-CM6A model in the CMIP6 experiment. We propose a novel causal inference framework that focuses on differences in return levels derived from annual temperature maxima between the factual and counterfactual worlds. While jointly modeling the annual maxima from the two worlds using a bivariate generalized extreme value distribution, we model the spatially-varying coefficients using a latent Gaussian framework. Specifically, given that the data are available over a $1^\circ \times 1^\circ$ grid, we employ the multivariate intrinsic conditional autoregressive model for the latent layer in the proposed hierarchical model, ensuring proper posterior distributions. We implement a recently developed highly-efficient approximate Bayesian inference technique, `Max-and-smooth', that uses a Laplace approximation of the likelihood and then performs Gibbs sampling based on the approximate posterior. The results include posterior estimates of the causal effect of anthropogenic forcing on high-temperature extremes, along with the trends in this effect, over the factual world. Furthermore, we estimate credible regions for a significant causal effect to facilitate hotspot detection across the mainland United States.
\end{abstract}


\section{Introduction}
\label{sec:introduction}

Identifying the various sources and their respective contributions to climate change is essential for formulating effective adaptation and mitigation strategies. The Sixth Assessment Report of the Intergovernmental Panel on Climate Change \citep[IPCC,][]{lee2023ipcc} provides comprehensive information on the current status and trends of climate change, the associated near- and long-term risks, and the progress of mitigation and adaptation policies. Since the pre-industrial era, human intervention has been a primary driver of climate change \citep{fischer2015anthropogenic}. Attributing anthropogenic forcing to the increased frequency of climate extremes, such as high-temperature and extreme precipitation events \citep{zhang2013attributing}, is central to Extreme Event Attribution \citep[EEA,][]{stott2016attribution} research. Numerous statistical methods have been developed to address the objectives of EEA; \cite{naveau2020statistical} provide an extensive review of these approaches. The principal aim of EEA is to compare the probability of an extreme event occurring in the factual world with its probability in a counterfactual world, defined as a world without anthropogenic forcing. The initial step involves designing the modeling framework for univariate extreme climate indices in both factual and counterfactual scenarios. For univariate extremes, whether blockwise maxima or threshold exceedances, established modeling techniques and inference methods are available in the extreme value theory literature \citep{coles2001introduction, davison2015statistics}. In the context of joint modeling of two blockwise maxima components and their dependence, a substantial body of literature addresses statistical methodologies for bivariate extremes \citep{tawn1988bivariate}. For comparisons between actual and hypothetical worlds, a widely adopted approach is to incorporate extreme-event definitions, using appropriate threshold selection, into causality metrics such as the Fraction of Attributable Risk \citep[FAR,][]{kiriliouk2020climate}. These causality metrics are equivalent to those in causal counterfactual theory \citep{hannart2016causal}, which are expressed as the probability of necessary causation (PN), the probability of sufficient causation (PS), and the probability of necessary and sufficient causation (PNS). However, integrating standard or spatial causal inference frameworks and methods to address EEA challenges remains largely unexplored.

The adoption of the causal inference paradigm, in conjunction with the potential outcome framework \citep{rubin1978bayesian}, has increased in the environmental and epidemiological domains. Accurate identification of causal effects necessitates rigorous application of methods that address spatial confounding \citep{gilbert2021causal}. Additionally, spatial interference introduces further complexity by violating assumptions such as the Stable Unit Treatment Value Assumption (SUTVA), an issue addressed in several recent studies \citep{papadogeorgou2023spatial, giffin2023generalized}. Various methods have been developed for valid causal inference in geospatially referenced settings; \cite{reich2021review} provide a comprehensive discussion of these techniques in the spatial causal inference literature. \cite{larsen2022spatial} apply the spatial causal framework to efficiently analyze the causal impact of wildland fire-contributed PM$_{2.5}$, incorporating the spillover effect. However, the implementation of practical methods for causal inference in the context of spatial extremes remains underexplored.

\cite{katzfuss2017bayesian} adopt a different approach based on Bayesian hierarchical modeling rather than the usual strategy, namely, nonparametric estimation of various causality metrics, such as PN, PS, and PNS, for extreme event attribution problems. The most important feature of Bayesian hierarchical modeling is the improved uncertainty assessment of inferred quantities, achieved by proposing an appropriate model for the latent layer and assigning priors to the hyperparameters that govern the latent variables. Several classes of Bayesian hierarchical models, including Bayesian Latent Gaussian models (LGMs), are well developed and support efficient spatial extremes modeling \citep{sang2009hierarchical, cooley2010spatial, reich2019spatial, johannesson2022approximate, hazra2023bayesian}. In a gridded data setting, we need appropriate models to capture all the important features, such as proper joint distribution specification and identification of dependence features within the same areal data. \cite{besag1974spatial} establish the foundation of areal data modeling by proposing the conditional autoregressive (CAR) model, which has been used extensively in the past decades. Besides, several extensions of the CAR model have been proposed in the literature. \cite{besag1995conditional} propose the intrinsic autoregressive model, which offers advantages over the usual autoregressive model, and demonstrate this by applying both models to agricultural experiments. \cite{lavine2012rigorous} provide a detailed description of the intrinsic conditional autoregressive (ICAR) model. From an application perspective, while \cite{cressie2011statistics} consider the ICAR model as a prior distribution in the context of a Bayesian analysis for ecological models in the spatiotemporal context, \cite{banerjee2003hierarchical} use the ICAR prior for modeling spatial rainfall variability. Extending the regular CAR model to dependence modeling settings, \cite{bhowmik2025bayesian} use the ICAR prior for spatially-varying coefficients in their proposed CAR copula model. In the multivariate setup, \cite{gelfand2003proper} extend the univariate CAR model to the Multivariate CAR (MCAR) model and apply it to the spatial modeling of child growth datasets.

In this paper, we propose a unified causal quantity, constructed using the potential outcome framework \citep{rubin1978bayesian}, based on return levels of annual maximum temperature, to quantify the effect of anthropogenic forcing on high-temperature extremes. As here our cause of interest is the presence of human intervention, we yield two distinct potential outcomes. For them, we use numerical climate model output from two scenarios: one with human forcing and the other with only natural forcing. We use the Generalized Extreme Value (GEV) distribution \citep{davison2015statistics} to model annual temperature maxima in the observation layer. The addition of nonstationarity and the reparameterization of some distributional parameters make our model more reasonable; we refer to LGMs. The entire study is conducted in the areal data setup because climate model outputs are available in gridded form. Given the extensive flexibility of the ICAR prior for spatially-varying coefficients in areal data, we explore its multivariate extension (MICAR) as a prior for the latent layer. Given the relevance of the latent-layer modeling approach, we examine available covariates and select a subset to model the mean component of the LGM prior. We employ a recently developed, highly accurate, and efficient approximate Bayesian inference method called ``Max-and-smooth'' \citep{hrafnkelsson2021max}, which applies a Laplace approximation to the likelihood and subsequently performs Gibbs sampling using the resulting approximate posterior. This approach yields posterior estimates of the causal impact of anthropogenic forcing on high-temperature extremes, as well as temporal trends in this effect within the factual world.

In addition to modeling data from both factual and counterfactual worlds using latent Gaussian models and inferring the causal effect of anthropogenic forcing, it is equally important, particularly in environmental and ecological applications \citep{patil2010digital}, to identify regions exhibiting extreme values over a spatial domain. This study focuses on detecting risk-prone areas where the estimated causal effect surpasses a specified threshold. In climate science, numerous studies have identified regions experiencing significant changes using indices such as temperature \citep{furrer2007spatial}, precipitation \citep{sain2011spatial}, and vegetation \citep{bolin2009fast}. Rather than conducting independent tests at each grid cell \citep{eklundh2003vegetation}, recent methods account for spatial dependence and multiple testing. These approaches characterize the joint occurrence of extreme events through threshold exceedance regions. Notable developments include \cite{cressie2005geostatistical}, \cite{french2013spatio}, and \cite{bolin2015excursion}. For hotspot identification, these approaches typically define regions where the joint probability of exceeding a threshold attains a specified level. For instance, \cite{french2013spatio} employ a frequentist hypothesis-testing framework to construct confidence regions with prescribed coverage. In contrast, \cite{french2016credible} introduce a Bayesian framework for constructing credible regions for exceedance sets using sampling-based methods. Extending this line of research, \cite{hazra2021estimating} estimate hotspots in the Red Sea using sea surface temperature data. In the present study, after estimating the causal effect, we adopt the Bayesian credible region estimation approach of \cite{french2016credible}. This method is chosen to identify hotspots of significant causal effect of anthropogenic forcing, due to its computational simplicity and effective use of posterior samples from the joint predictive distribution.

The rest of the paper is structured as follows. Section~\ref{sec:background} summarizes the necessary background about univariate extreme modeling and the bivariate Husler-Reiss distribution. In Section \ref{sec:EDA}, we present the dataset comprising climate model outputs from the IPSL-CM6A model and the covariates used for statistical analysis, along with some exploratory analyses.  Section \ref{sec:methodology} discusses our proposed framework, including the necessary assumptions and the model description. In Section \ref{sec:inference_hotspot}, we describe the approximate Bayesian inference technique Max-and-Smooth, and a hotspot estimation algorithm. In Section \ref{sec:data_application}, we apply the proposed methodology to the dataset and discuss the results and key findings. Finally, we finish with some concluding remarks and an outlook on future directions in Section \ref{sec:discussion}.
\section{Background}\label{sec:background}

\subsection{Univariate extreme value theory}
\label{subsec:univariate_evt}

The extreme value theory literature primarily focuses on two methodologies for modeling univariate extremes. One approach is the block maxima method, which typically involves choosing a suitably large block size and analyzing the block maxima using their asymptotic distribution as the block size tends to infinity. Generally, from an application perspective, practitioners prefer annual, monthly, or weekly maxima, depending on the problem at hand. The other one is the peak-over-threshold approach, where the usual analysis involves choosing an appropriate high threshold and modeling threshold exceedances using a generalized Pareto distribution. Further details are in \cite{coles2001introduction} and \cite{davison2015statistics}. While both approaches have their own pros and cons, we stick to the block maxima approach here. Let $\{X_n \}_{n \geq 1}$ be a sequence of independent random variables having a common distribution $F(\cdot)$, and we define the block maxima by $M_n = \text{max}\{X_1,\ldots, X_n\}$. Following Fisher-Tippett theorem \citep{fisher1928limiting}, if there exist sequences of normalizing constants $\{a_n > 0\}$ and $\{b_n \in \mathbb{R}\}$ such that the rescaled maximum $M_n^{*} =(M_n-b_n )/ a_n$ has a nondegenerate limiting distribution as $n \rightarrow \infty$, then the limiting distribution can be either Gumbel, or Fr\'echet, or reversed Weibull distributions; the three forms can be merged to yield the Generalized Extreme Value (GEV) distribution, having the distribution function
\begin{equation}\label{eq:gev_exp}
    F_{\mathrm{GEV}}(z;\mu,\sigma,\xi)=
      \begin{aligned}
      \begin{cases}
            \exp \biggl\{ -\biggr[1 + \xi\Bigl(\dfrac{z-\mu}{\sigma}\Bigl)\biggr]_{+}^{-\frac{1}{\xi}}\biggl\}, & \xi \neq0, \\
            \exp\biggl\{-\exp\Bigr[-\dfrac{(z-\mu)}{\sigma}\Bigr]\biggl\}, &     \xi = 0,
        \end{cases}
       \end{aligned} 
    \end{equation}
where $m_+=\max\{m,0\}$. Specifying the parameters of the GEV distribution, $\mu \in \mathbb{R}$ denotes the location parameter, $\sigma > 0$ denotes the scale parameter, and $\xi \in \mathbb{R}$ denotes the shape parameter that signifies the weight of the upper tail of the density. Support of $F_{\mathrm{GEV}}$ is $\mathbb{R}$ if $\xi = 0$, $(-\infty, \mu-\sigma/\xi)$ if $\xi < 0$, and $(\mu - \sigma/\xi, \infty)$ if $\xi > 0$. 

A primary objective of extreme value analysis is to extrapolate from the observed maximum over a given time period. Suppose $Z$ is the block maxima of variables of interest, such that $Z \sim \text{GEV} (\mu,\sigma,\xi)$. We are interested in estimating the return level $z_p$, which is defined as $\mathrm{Pr}(Z> z_p) = p$. Using \eqref{eq:gev_exp} and assuming $1+[\xi(z_p-\mu)/\sigma] >0$, $z_p$ can be represented as 
\begin{equation}\label{eq:rtr_lvl}
    z_p =
    \begin{cases}
        \mu + \dfrac{\sigma}{\xi}\Bigl[\{-\log(1-p)\}^{-\xi}-1\Bigl], & \xi \neq0, \\
        \mu - \sigma[\log\{-\log(1-p)\}], &     \xi = 0.
    \end{cases}
\end{equation}
Here, $p\in (0,1)$ denotes the excess probability, and $z_p$ denotes the return level associated with the return period $1/p$.
\subsection{Bivariate H\"usler-Reiss distribution}
\label{subsec:bivariate_evt}

We first present a framework that incorporates the asymptotic convergence of bivariate component-wise maxima for the analysis of bivariate extremes. Suppose $\{(X_n, Y_n)\}_{n \geq 1}$ is a sequence of bivariate random vectors having the common distribution function $F(\cdot,\cdot)$. We define the bivariate block maxima using a similar approach to that for univariate block maxima; let $M_{1n}=\text{max}\{X_1,\ldots, X_n\}$ and $M_{2n}=\text{max}\{Y_1,\ldots, Y_n\}$ be the respective component-wise block maxima. We define the bivariate component-wise block maxima as $\bm{M}_n=(M_{1n},M_{2n})^\top$. Similar to the univariate case, if there exist two sequences of normalizing constants $\{a_{in} >0\}$ and $\{b_{in} \in \mathbb{R}\}$, for $i=1,2$, such that 
\begin{equation}
    \mathrm{P}\{(M_{1n}-b_{1n})/a_{1n} \leq x, (M_{2n}-b_{2n})/a_{2n} \leq y\} = F^n(a_{1n}x+b_{1n}, a_{2n}y+b_{2n}) \rightarrow G(x,y),
\end{equation}
as $n \rightarrow \infty$, where $G$ is a nondegenerate bivariate probability distribution, then $G$ is necessarily a bivariate GEV distribution function; we denote it by $F_{\mathrm{BGEV}}$. Considering $M_{1n}\sim\text{GEV}(\mu_{1},\sigma_{1},\xi_{1}) $ and $M_{2n}\sim\text{GEV}(\mu_{2},\sigma_{2},\xi_{2})$ in the limiting sense, the CDF $F_{\mathrm{BGEV}}$ can be expressed as %
\begin{equation}
\nonumber F_{\mathrm{BGEV}}(x,y) = \exp\{-V(\Tilde{x},\Tilde{y})\},
\end{equation}
where $\Tilde{x} = [ 1 + \xi_{1}(x-\mu_{1})/\sigma_{1}]_{+}^{-1/\xi_{1}} $ and $\Tilde{y}= [ 1 + \xi_{2}(y-\mu_{2})/\sigma_{2}]_{+}^{-1/\xi_{2}}$. Here, $V$ is called exponent measure of $F_{\mathrm{BGEV}}$. Further details are in \cite{coles2001introduction} and \cite{tawn1988bivariate}.

\cite{sibuya1960bivariate} prove the asymptotic independence between marginal extremes by considering a sequence of independent and identically distributed bivariate normal random vectors; however, such a theoretical property is unrealistic in the case of some real-world situations. Considering Pearson's correlation between $X_n$ and $Y_n$ to be $\rho_n$, i.e., dependent on $n$, and $\lim_{n \rightarrow \infty} (1 - \rho_n) \log(n) = \lambda^{-2}$, \cite{husler1989maxima} propose a form of the bivariate GEV distribution, namely the bivariate H\"usler-Reiss (BHR) distribution, that correctly quantifies asymptotic independence. There are several equivalent representations of the BHR distribution via transformations of the components to standard Gumbel margins or standard Fr\'echet margins. Here, we stick to the Fr\'echet margin as in \cite{stephenson2018evd}. The symmetrical BHR distribution function is represented as
\begin{equation}\label{eq:bvhr}
    F_{\mathrm{BHR}}(x,y; \bm{\theta}_{\mathrm{BHR}}) = \exp\Bigg[-\Tilde{x}\Phi\biggl(\frac{1}{\lambda} + \frac{\lambda}{2}\log\biggl(\frac{\Tilde{x}}{\Tilde{y}}\biggl)\biggl) - \Tilde{y}\Phi\biggl(\frac{1}{\lambda} + \frac{\lambda}{2}\log\biggl(\frac{\Tilde{y}}{\Tilde{x}}\biggl)\biggl)\Bigg],
\end{equation}
where $\Tilde{x} = \{ 1 + \xi_{1}(x-\mu_{1})/\sigma_{1}\}^{-1/\xi_{1}} $, $\Tilde{y}= \{ 1 + \xi_{2}(y-\mu_{2})/\sigma_{2}\}^{-1/\xi_{2}}$, $\lambda > 0$ denotes the strength of dependence, and $\bm{\theta}_{\mathrm{BHR}} = (\mu_1,\sigma_1,\xi_1, \mu_2,\sigma_2,\xi_2, \lambda)^\top$.
\subsection{Bayesian latent Gaussian model}
\label{subsec:bayesian_lgm}

Incorporation of prior scientific information on climatological and other geophysical processes, and the coherent exposition of the uncertainties related to these processes during modeling, are often achieved in the literature through Bayesian hierarchical models, where a suitable model for the observations in the data layer induces the latent variables. The hyperparameters, generated by assigning a prior of choice to the latent layer, characterize the latent model and provide insights into features such as variability and dependencies. In general, reflecting limited prior knowledge about hyperparameters, weakly-informative priors are often preferred in practice. Suppose $\bm{y},\bm{\theta}$, and $\bm{\phi}$ represent the observations, the latent variables, and the hyperparameters of the latent variables, respectively. The joint density of $(\mathbf{y},\boldsymbol{\theta},\boldsymbol{\phi})$ is then expressed as 
\begin{equation}
\nonumber \pi(\bm{y},\bm{\theta},\bm{\phi})=\pi(\bm{y}|\bm{\theta}) \pi(\bm{\theta}|\bm{\phi}) \pi(\bm{\phi}),
\end{equation}
where $\pi(\bm{y}|\bm{\theta})$ denotes the conditional density of observations given latent variables, $\pi(\bm{\theta}|\bm{\phi})$ denotes the conditional density of latent variables given their hyperparameters, and $\pi(\bm{\phi})$ is the marginal density of hyperparameters. Here, the hierarchical nature of the model gives rise to the conditional densities. To a large extent, the primary objective is to draw posterior inference corresponding to the latent variables. For drawing posterior inference corresponding to the latent variables as well as their hyperparameters, we rely on the joint posterior density represented as
    \begin{equation} \label{eq:posterior}
        \pi(\bm{\theta},\bm{\phi}|\bm{y})
         = \frac{\pi(\bm{y},\bm{\theta},\bm{\phi})}{\pi(\bm{y})}
         \varpropto \pi(\bm{\phi}) \pi(\bm{\theta}|\bm{\phi}) \pi(\bm{y}|\bm{\theta}).
    \end{equation}

Selecting the distribution of latent variables to be Gaussian yields a broad subclass of Bayesian hierarchical models known as Bayesian latent Gaussian models (LGMs); further details are in \cite{hrafnkelsson2023statistical}. Here, the above posterior density can be rewritten as
\begin{equation}
\nonumber    \pi(\bm{\theta},\bm{\phi}|\bm{y}) \varpropto \pi(\bm{\phi}) \mathcal{N}(\bm{\theta}|\bm{\mu_{\phi}},\bm{\Sigma}_{\bm{\phi}})\pi(\bm{y}|\bm{\theta}),
\end{equation}
where $\mathcal{N}(\cdot | \bm{\mu_{\phi}},\bm{\Sigma}_{\bm{\phi}})$ denotes a multivariate Gaussian density with mean component $\bm{\mu_{\phi}}$ and covariance matrix $\bm\Sigma_{\bm{\phi}}$. Here, $\bm{\mu_{\phi}}$ and $\bm\Sigma_{\bm{\phi}}$ are user-defined functions of $\bm{\phi}$. Using the Gaussian distribution to model the observations, i.e., $\pi(\bm{y}|\bm{\theta})$, simplifies posterior inference and leads to a special class of models known as Gaussian-Gaussian models. When modeling extreme climate indices in the data layer, we often use heavy-tailed non-Gaussian distributions, e.g., the univariate or bivariate GEV distributions described in Sections \ref{subsec:univariate_evt} and \ref{subsec:bivariate_evt}. Proposing a non-Gaussian likelihood for the data layer, combined with a Gaussian prior for the latent layer, does not lead to closed-form expressions of the posteriors or full conditional posteriors, and increases the complexity of posterior inference. From a computational viewpoint, in Bayesian computation for complex spatial extremes models, a key objective is to develop a faster method for posterior inference that mitigates the computational burden due to high dimensionality in both the observed data and the latent variables. The approximate Bayesian inference approach `Max-and-Smooth' \citep{hrafnkelsson2021max} is applicable when the data layer is modeled using an additive-Gaussian process. \cite{johannesson2022approximate} extend the Max-and-Smooth approach to a non-Gaussian-Gaussian setting, where the distribution for the data layer is univariate GEV and for the latent layer is Gaussian. While the data distribution is non-Gaussian, a Laplace approximation of the data likelihood yields a Gaussian-Gaussian pseudo-model, removing a layer of complexity in posterior computation. The appropriate use of this approach in our setup is described in Section \ref{sec:inference_hotspot}.


\section{Dataset and exploratory analysis}\label{sec:EDA}

In our analysis, we curate a dataset from the French global climate model IPSL-CM6A \citep{boucher2020presentation}, which is part of the sixth phase of the Coupled Model Intercomparison Project (CMIP6). The climate model outputs consist of daily temperature series. Here, our domain of interest is the mainland United States, divided into $1^\circ \times 1^\circ$ grid cells, for a total of $250$ cells. We consider the time period from $1\text{st}$ January, 1850, to $31\text{st}$ December, 2014, which totals $165$ years. The factual world corresponds to the historical run with the anthropogenic forcing (HIST), and the counterfactual corresponds to the run with only natural forcing (NAT). We preprocess the climate model outputs, in the form of daily temperatures, to obtain annual temperature maxima, yielding $165$ maxima per grid cell.

Overall, for each grid cell-year pair, we have a bivariate vector of annual temperature maxima, with components corresponding to the factual and counterfactual worlds. Here, the climate model simulations for the two worlds differ only in the presence or absence of anthropogenic forcing, and other potential forcings are common. As a result, the components of the above-mentioned bivariate vector are likely to be dependent, and we assume it to follow a bivariate H\"usler-Reiss (BHR) distribution. Given the long data availability period (165 years), it is reasonable to include a temporal trend component in the marginal distributions, and it is common to include one in the location parameters. Further, considering parsimony, we assume that the trends are linear, the scale and shape parameters are invariant across the factual and counterfactual worlds and across years; still, we allow them to vary across space. The BHR dependence parameter is also assumed to vary across space but to be temporally invariant. Finally, the joint distribution for each grid cell-year pair is specified through \eqref{eq:pot} and \eqref{eq:bivariate_pot}; While we discuss the model specifications in more detail in Section \ref{sec:methodology}, the joint distribution is parameterized by $\{\alpha_0(g), \alpha_1(g), \beta_0(g), \beta_1(g), \sigma^*(g), \psi(g), \lambda^*(g)\}$ for grid cell $g$, where the first two elements represent the intercept and slope terms in the location parameter for counterfactual world, and the third and the fourth entries represent the same for factual world. The fifth entry denotes the common scale parameter, but on a logarithmic scale. The sixth entry is a specialized transformation of the common shape parameter explained in \eqref{eq:xi2psi}. The seventh entry denotes the strength of dependence of the BHR distribution in \eqref{eq:bvhr} for grid cell $g$, but on a logarithmic scale.

For each grid $g$ over the spatial domain of interest, using the bivariate responses from all years, we first obtain the maximum likelihood estimates (MLEs) of $\{\alpha_0(g), \alpha_1(g), \beta_0(g), \beta_1(g), \newline \sigma^*(g), \psi(g), \lambda^*(g)\}$, based on the BHR likelihood. The spatial maps of the MLEs are provided in the supplementary material; they show a clear spatially-varying pattern, with nearby regions having similar values. In our proposed LGM, we aim to specify a joint Gaussian distribution for the latent variables, i.e., the spatially-varying parameters of the BHR distributions. The MLEs obtained above are reasonable naive estimates of the latent variables; thus, we explore the potential features of their joint distribution, i.e., the mean and covariance structures, using the MLEs. To model the mean structure, we include longitude, latitude, mean elevation, and open sea distance for each grid as covariates. The scatter plots of the available covariates versus the MLEs are presented in the supplementary material; a linear dependence structure between the responses and the covariates appears reasonable for most cases. We thus first regress the MLEs of each latent variable on the chosen covariates using a multiple linear regression model. After fitting the linear model, the estimated regression coefficients and their respective standard errors are shown in Table \ref{est_reg_lm}. The statistical significance of the regression coefficient estimates indicates the extent to which the chosen covariates explain the variability of the latent variables.

\begin{figure}[t]
    \centering
\includegraphics[height=0.4\linewidth]{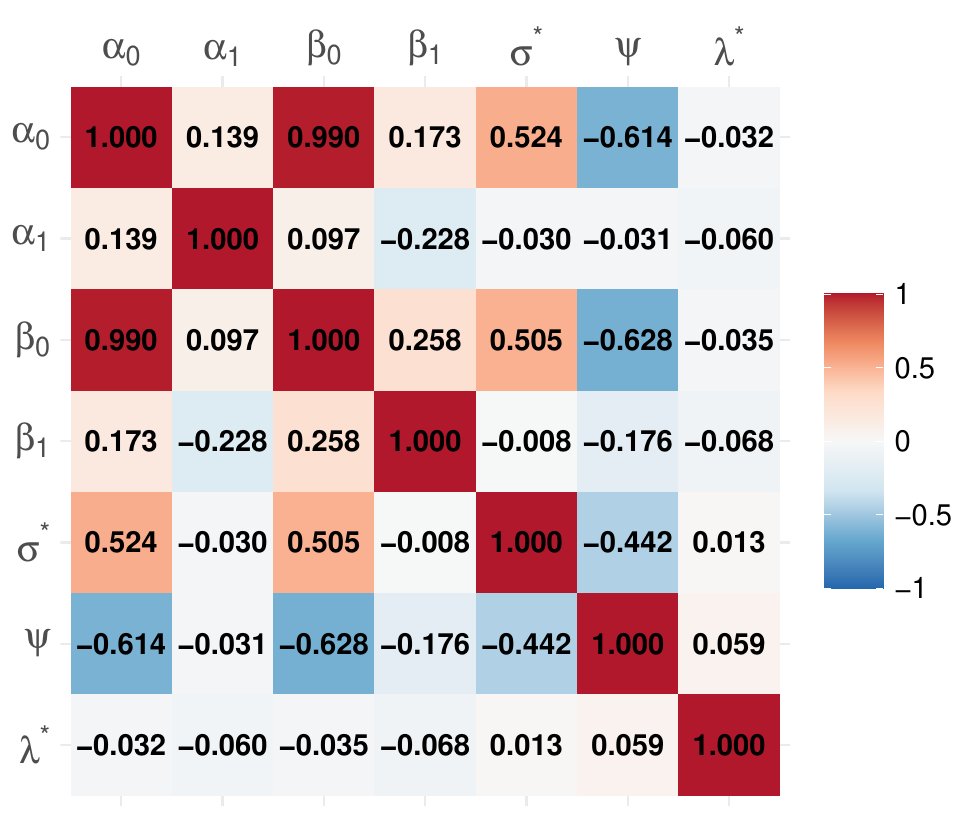}
 \includegraphics[height=0.4\linewidth]{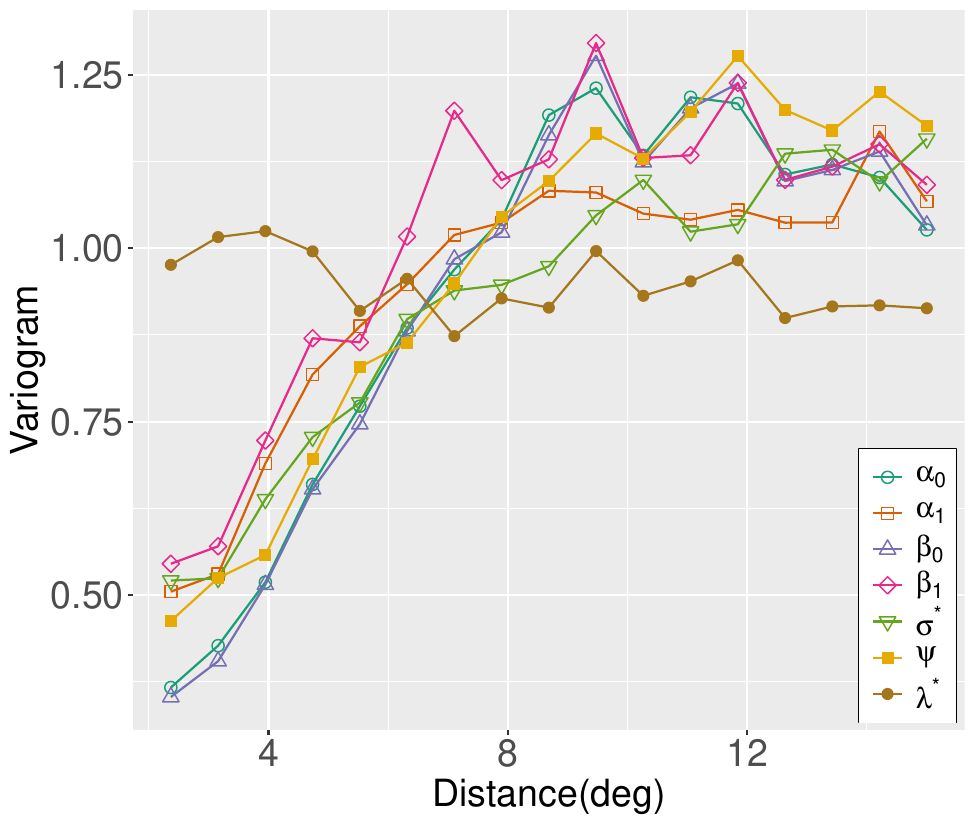}
\caption{Sample correlation matrix based on residuals corresponding to all seven latent parameters, obtained after fitting the linear model (left panel) and variogram plot of residuals corresponding to all seven latent parameters, obtained after fitting separate multiple linear regression models (right panel). Here, we consider Euclidean distances between grid-cell centroids.}
\label{res_eda}
\end{figure}

\begin{table}[h]
    \centering
     \caption{Estimated regression coefficients, with respective standard errors in brackets, for each latent variable corresponding to chosen covariates. Here, Int denotes the intercept, Lon denotes the Longitude, Lat denotes the latitude, ME denotes the mean elevation, and OS denotes the open sea distance (from the centroids).}
    \setlength{\tabcolsep}{0.3pt}
    \begin{tabular}{|l|c|c|c|c|c|c|c|}
    \multicolumn{1}{c}{}&
    $\alpha_0$ &  
    $\alpha_1$ &
    $\beta_0$ &
    $\beta_1$ &
    $\sigma^*$ &
    $\psi$  &
    \multicolumn{1}{c}{$\lambda^*$} \\
    \hline   
    Int    &310(1.07)   &-0.13(0.06)  &311.5(1.07)   &0.09(0.08)   &-1.01(0.08)  &-0.05(0.04)  &0.17(0.21) \\
    Lon   &-71.61(9.22)  &0.80(0.50) &-64.54(9.21) &-0.48(0.70) &-8.46(0.75)  &3.31(0.37)  &4.93(1.85) \\
    Lat  &-420.90(21.45) &8.65(1.18)  &-432.30(21.44)  &7.35(1.65)  &9.97(1.75)  &6.05(0.86)  &10.75(4.32)\\
    ME    &-5.40(0.18) &0.03(0.01)  &-5.19(0.18)  &0.11(0.01) &-0.21(0.01)  &-0.01(0.02)  &0.07(0.04)\\
    OS   &4.10(0.23)  &0.06(0.01) &3.66(0.23)  &-0.28(0.01)  &0.28(0.02)  &-0.08(0.01)  &-0.21(0.05)\\
    \hline
    \multicolumn{8}{l}{\footnotesize Here, except for the elements of the first row, elements of all the remaining rows are multiplied by $10^{3}$.}
    \end{tabular}
    \label{est_reg_lm}
\end{table}
We further explore the choices for the dependence structure of the LGM after removing covariate effects, i.e., using the residuals from the multiple linear regression model fit explained above. We first construct the sample correlation matrix for all seven parameter types, ignoring spatial structure, and present it in Figure \ref{res_eda}. The nonzero entries in the sample correlation matrix reflect the correlation and variability of latent parameters within a particular grid, underscoring the importance of considering a cross-component covariance matrix in the overall latent-layer model to capture intravariability within the grid. Further, we explore the strength of spatial dependence through empirical variograms \citep{cressie2015statistics} for all latent variables. Using the residuals obtained above and considering the Euclidean distance (in degrees) between the centroids of the grid cells, we estimate the variogram of residuals separately for each of the seven latent parameters. The estimated semivariances as a function of Euclidean distances are presented in Figure \ref{res_eda}. For almost all latent variables, the estimated semivariance values start at relatively low levels, jump to higher levels, and remain there. This behavior indicates the need to account for spatial correlation among the latent variables. A slightly different behavior is observed for the process $\lambda^*$; while treating this latent process differently from the others might be more appropriate, for convenience, we treat all seven latent processes similarly. Overall, choosing a spatially dependent prior structure is important here for uncertainty quantification via Bayesian inference. For this purpose, we choose the conditional autoregressive (CAR) model as the prior for the latent layer. Besides, the semivariance profiles for all latent processes except $\lambda^*$ showcase a similar spatial range and total variation. Considering these factors and parsimony, assuming a separable covariance structure across the space and components appears legitimate. Besides, a specific case of a CAR prior, namely the intrinsic CAR prior, is more parsimonious, with the spatial autocorrelation parameter fixed at one. This improper prior also allows spatial shrinkage, and the posterior distribution is proper. Thus, overall, a separable multivariate intrinsic CAR prior for the latent variables is suitable in our setup. 

While we explore the dependence structure of the latent processes, we estimate the grid-cell-specific parameters independently across grid cells, ignoring potential spatial dependence in the data layer. Given that the observations represent annual maxima, we explore the extremal dependence of the data layer. The extremal dependence between two random variables $Y_1$ and $Y_2$ is defined using the $\chi$ measure \citep{sibuya1960bivariate}, given by
\begin{equation}
 \nonumber   \chi= \lim_{u\to1}\chi(u)=\lim_{u\to1}\text{Pr}[Y_1>F_1^{-1}(u)|Y_2>F_2^{-1}(u)],
\end{equation}
 where $Y_1\sim F_1$ and $Y_2 \sim F_2$, respectively. Here, $\chi = 0$ indicates asymptotical independence and $\chi > 0$ implies asymptotical dependence. In a geospatial setting, assuming the extremal dependence to be isotropic in nature \citep{hazra2020multivariate}, it can be redefined as the extremal dependence between two locations, i.e.,
 \begin{equation}\label{eq:emp_chi}
     \chi(\Vert \bm{s} - \bm{s}' \Vert)= \lim_{u \to 1}\text{Pr}[Y(\bm{s}')>F_{Y(\bm{s}')}^{-1}(u)|Y(\bm{s})>F_{Y(\bm{s})}^{-1}(u)],
 \end{equation}
where $Y(\cdot)$ is a spatial stochastic process defined over a domain $\mathcal{D}$, with $Y(\bm{s}) \sim F_{Y(\bm{s})}$ for each $\bm{s} \in \mathcal{D}$ and $\Vert \bm{s} - \bm{s}' \Vert$ is the Euclidean or geodesic distance between $\bm{s}$ and $\bm{s}'$. Here, we use this extremal dependence measure to support the assumption that the observations are conditionally spatially independent in the tail, given the latent variables. After taking into account all temporal observations throughout the spatial domain during the observation period, we calculate the empirical $\chi$ measure, where we fix $u=0.95$ in \eqref{eq:emp_chi} instead of the limiting behavior. We calculate the empirical $\chi$ measure separately for both the counterfactual and factual worlds using observations from different runs of the numerical climate model IPSL-CM6A. The distance-versus-extremal-dependence profiles are presented in the supplementary material. Although strictly the range of values obtained by the empirical $\chi$ measure is not negligible, it is moderate. Besides, we study the empirical $\chi$-measure for $u=0.95$ due to only 165 temporal replications, and a sharp decay of empirical $\chi$-measure over higher levels of $u$, say $u=0.99$, is common in practice \citep{hazra2025efficient}. In this context, \cite{davison2012statistical} review different statistical modeling strategies for spatial extremes. Here, the authors discuss the effectiveness of latent variable modeling compared to the max-stable process, with the former emphasizing a better fit to the marginal distribution and the latter having an upper hand when the objective is to capture the joint distribution over the spatial domain efficiently. Here, our primary objective is to quantify the causal effect of anthropogenic forcing on temperature extremes, using marginal return levels. Hence, we prefer to use the LGM framework in an analogous manner to \cite{johannesson2022approximate}.

\section{Methodology}\label{sec:methodology}

Here, we propose a novel Bayesian hierarchical model with a data layer modeled using the bivariate H\"usler-Reiss (BHR) distribution, and a latent layer consisting of transformed spatially-varying parameters jointly modeled using multivariate spatial priors. We also discuss the proposed causal inference framework and the necessary assumptions for causal effect identification. Our framework is based on the potential outcomes model, widely used in causal inference.

\subsection{Data layer modeling specifications} \label{subsec:data_layer}

Let $Y_t(g_i)$ be the observed annual maximum temperature at the grid cell $g_i$, where $g_i \subset \mathcal{D} \subset\mathbb{R}^2$, $i \in \mathcal{I} = \{1,\cdots,250\}$, $t \in \mathcal{J}$ and $\mathcal{J}$ is the index set that contains all the indices corresponding to the years at which the temperature maxima are observed. The observations at a particular grid cell $g$ are assumed to be independent across years, a standard assumption in the literature. The observed maxima, i.e., $Y_t(g)$ at that grid cell $g$ and year $t$ is modeled using the non-stationary GEV distribution having the time-dependent location parameter $\mu_{t}(g)$, scale parameter $\sigma(g)$, and shape parameter $\xi(g)$, that is,
\begin{equation}\label{eq:gev}
    Y_t(g) \sim \text{GEV}(\mu_t(g),\sigma(g),\xi(g)), 
\end{equation}
where $\mu_t(g) \in \mathbb{R}$, $\sigma(g) >0$, $\xi(g) \in (-0.5,0.5)$, and for all the realizations $y_t(g)$, $1+\{\xi(g)(y_t(g)-\mu_t(g))/\sigma(g)\} > 0$. The choice of the interval for $\xi(g)$ allows the existence of the marginal variance and valid large sample theory results. We assume the data to be conditionally independent given the latent GEV parameters. Here, our main goal is to estimate the causal effect using marginal return-level estimates, which supports the feasibility of our conditional independence assumption. Instead of directly specifying Gaussian priors for spatially-varying parameters, we reparameterize the scale parameter $\sigma(g)$ and the shape parameter $\xi(g)$. The transformed parameters are used later in the latent layer when constructing the Bayesian hierarchical model. The transformed scale parameter is denoted by $\sigma^*(g) = \log[\sigma(g)] \in \mathbb{R}$. The transformed shape parameter is denoted by $\psi(g) = f(\xi(g))$; hereafter, we drop the argument $g$ for notational convenience in this subsection. The proper choice of $f$ informs the proposed transformation of the shape parameter. Similar to \cite{johannesson2022approximate}, the choice relies on three natural conditions: (1) the variance of the distribution (i.e., the GEV distribution) that is used to model our extreme observations should be finite, which leads to $\xi < 0.5$; (2) upper boundedness, that is, the upper bound of the GEV distribution should not be smaller than $\mu +2\sigma$, which leads to $\xi > -0.5$; (3) the transformation $f$ should be monotonic, such that for the values of $\xi$ around $\xi = -0.25$, $\psi$ is approximately equal to $\xi$. In contrast to the modeling of precipitation extremes, where the estimated shape parameter usually takes values around zero \citep{cooley2010spatial}, while modeling the annual temperature maxima, the mass of the estimated shape parameter is usually concentrated around significantly negative values, mostly around $-0.25$ \citep{cooley2009extreme, huang2016estimating, hogan2019representation}. In a manner analogous to \cite{johannesson2022approximate}, which considers values around zero to be approximately constant under their transformation scheme, we treat values around $-0.25$ as approximately constant under the proposed transformation method. An equivalent mathematical interpretation of the condition is that $\psi = f(\xi) \approx \xi$ when $\xi$ is close to $-0.25$. Hence,
\begin{equation}\nonumber
    \frac{df(\xi)}{d\xi} =f'(-0.25)=1, \space \space f(-0.25)=-0.25.
\end{equation}
All the restrictions on the shape parameter through the conditions motivate us to come up with the following unified equation, which reflects the transformation, that is, 
\begin{equation}\nonumber
    \psi = f(\xi) =a_\psi +b_\psi \log\Bigg(\frac{(\xi +1/2)^{c_{\psi}}}{1-(\xi +1/2)^{c_{\psi}}}\Bigg),
\end{equation}
where $c_\psi = 0.005$,
    \begin{align}\nonumber
        b_\psi 
        & = 0.25 c_\psi^{-1}\{1-(0.25)^{c_{\psi}}\}= 0.34538, \\ \nonumber
        a_\psi
        & = -0.25 - b_\psi\biggl\{\log \biggl(\frac{(0.25)^{c_{\psi}}}{1-(0.25)^{c_{\psi}}}\biggl)\biggl\}= -1.96589. \nonumber
    \end{align}
Moreover, the inverse of the transformation is,
\begin{equation}\label{eq:xi2psi}
    \xi = f^{-1}(\psi)=g(\psi)= \Bigg[1+\exp\biggl\{-\biggl(\frac{\psi-a_\psi}{b_\psi}\biggl)\biggl\}\Bigg]^{-1/c_\psi} - \frac{1}{2}.
\end{equation}

\begin{figure}[t]
  \centering 
  \includegraphics[height=0.3\linewidth]{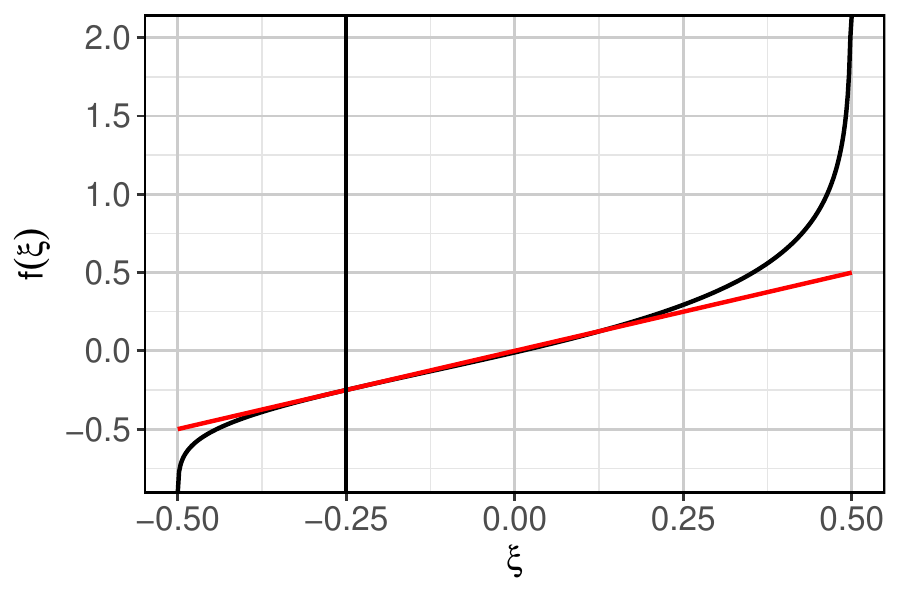}
  \includegraphics[height=0.3\linewidth]{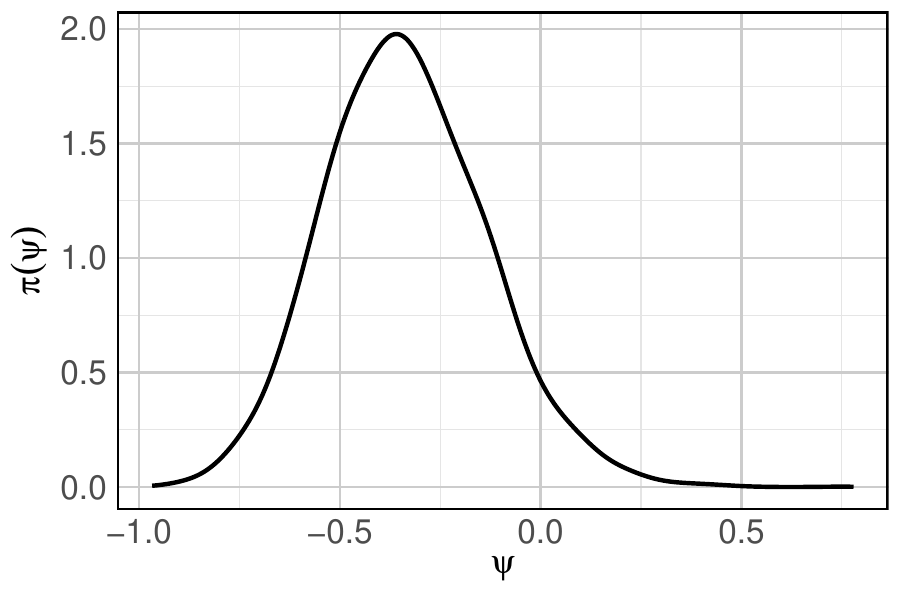}
  \caption{ The curve corresponding to the transformation of the shape parameter, $\psi= f(\xi)$, along with the reference line with slope one and intercept zero (left panel) and
  the prior density for the transformed shape parameter $\psi$ (right panel).}
  \label{fig:prior.psi}
\end{figure}

We adopt the approach of \cite{martins2000generalized}, namely the utilization of a generalized likelihood function to draw inference about the parameters of the GEV distribution. Given the common irregular behavior of maximum likelihood estimates when samples are drawn from a specified range of parameters, \cite{martins2000generalized} propose an alternate generalized likelihood estimation approach. More specifically, the generalized likelihood function is the modified likelihood, obtained by combining the likelihood of an extreme value distribution with a prior density that encapsulates all known behavior of $\xi$. \cite{martins2000generalized} use a shifted beta density having the support $(-0.5,0.5)$ with mean $0.10$ and standard deviation $0.122$ for flood data. Later, \cite{cooley2010spatial} use this stabilized approach by including the proper prior density of $\xi$ for modeling extreme precipitation from the Regional Climate Model (RCM). Similarly, \cite{johannesson2022approximate} refine this approach within a Bayesian inference framework while analyzing annual flood frequency data across the United Kingdom. We claim that $\xi$ should lie on the interval $(-0.5,0.5)$ with the help of conditions imposed on $\xi$, which should be feasible in most environmental applications. To a certain extent, we adapt the same argument as \cite{johannesson2022approximate} while choosing the prior density for $\xi$, and refine their choosing process according to the nature of the dataset used in the analysis. While fitting the GEV distribution to the observed annual maximum temperature data, we often observe that the estimated shape parameter is negative rather than positive \citep{hasan2013modeling,hogan2019representation}. This nature of the estimated shape parameter also implies that, on any particular day of a year, the maximum temperature is bounded above. Mathematically, in our case, the prior probability of $\xi$ taking values in $(-0.5,0)$ is higher than that in $(0,0.5)$, as we deal with the annual maximum temperature dataset. Accordingly, it is natural to consider a higher prior mass within $(-0.5,0)$ while selecting the prior density for $\xi$. Thus, we leverage an asymmetric beta prior density with shape parameters 1 and 4, respectively, shifted to the interval $(-0.5,0.5)$ for each $\xi(g)$. Subsequently, the beta prior density for each $\xi(g)$ is transformed to $\psi(g)$ using the Jacobian. The transformation $f$ and the final prior for $\psi(g)$ to be considered in the generalized maximum likelihood estimation in the first stage are presented in Figure \ref{fig:prior.psi}.

\subsection{Causal inference framework}
\label{subsec:causal_inference}

To estimate the effect of anthropogenic forcing on high-temperature extreme events, that is, high temperature extremes would not have occurred in the absence of anthropogenic forcing, we use the potential outcome framework \citep{neyman1923applications, rubin1978bayesian} as the primary building block. The first step towards applying the potential outcome framework \citep{rubin1978bayesian} in the realm of causal inference methodologies is to consider the treatment variable, which is the presence of anthropogenic forcing. Because the treatment variable is binary, there are two potential outcomes. One potential outcome is denoted by $Y_t(g,1)$, which is obtained in the presence of anthropogenic forcing, that is, the factual world. While the other one is denoted by $Y_t(g,0)$ and the only difference is the absence of anthropogenic forcing, which is obtained in the counterfactual world. Using \eqref{eq:gev}, both potential outcomes are modeled as
\begin{equation}\label{eq:pot}
    \begin{aligned}
        Y_t(g,0) |\alpha_0(g),\alpha_1(g),\sigma(g),\xi(g) \overset{\mathrm{ind}}{\sim} \text{GEV}(\alpha_0(g)+\alpha_1(g)t^*,\sigma(g),\xi(g)), \\
        Y_t(g,1) |\beta_0(g),\beta_1(g),\sigma(g),\xi(g) \overset{\mathrm{ind}}{\sim} \text{GEV}(\beta_0(g)+\beta_1(g)t^*,\sigma(g),\xi(g)),
    \end{aligned}
\end{equation}
where $\alpha_0(g),\alpha_1(g)$ are the location intercept and the time intercept parameter for grid $g$ in the counterfactual world. Here, $\overset{\mathrm{ind}}{\sim}$ denotes the conditional independence of the random variables across their indexes. Likewise, $\beta_0(g)$ and $\beta_1(g)$ signify the same thing for grid $g$ in the factual world. Here, $t^*$ is defined as $t^*=(\mathrm{year}_t-m_{\mathcal{T}})/sd_{\mathcal{T}}$, where $m_{\mathcal{T}}$ implies the mean of $\mathcal{T}$, $sd_{\mathcal{T}}$ corresponds to the standard deviation of $\mathcal{T}$, for $\mathcal{T}=\{1850,\cdots,2014\}$, the entire observational period and $\mathrm{year}_t$ is the year corresponding to index $t$. This temporal standardization setting is common in practice. The time-invariant parameters, $\sigma(g)$ and $\xi(g)$, are the scale and transformed shape parameters at grid $g$. Here, we assume that the scale and shape parameters are the same in both the counterfactual and factual worlds, based on our exploratory data analysis. When designing numerical climate models for both historical and natural experiments, scientists consider some common parameters whose effects must be accounted for. For common climate drivers, such as large-scale circulation patterns including the El Niño-Southern Oscillation (ENSO), that influence high-temperature extremes in both the counterfactual and factual worlds, it is reasonable to construct a bivariate extreme vector using potential outcomes, with one component from the counterfactual world and the other from the factual world. We use the BHR distribution in \eqref{eq:bvhr} to model the bivariate extreme data vector $(Y_t(g,0), Y_t(g,1))^\top$. We represent the bivariate model as
\begin{equation}\label{eq:bivariate_pot}
    (Y_t(g,0),Y_t(g,1))^\top |\bm{\eta}(g) \overset{\mathrm{ind}}{\sim} \text{BHR} (\bm{\eta}(g)),
\end{equation}
where $\bm{\eta}(g) = (\alpha_0(g), \alpha_1(g), \beta_0(g), \beta_1(g), \sigma^*(g), \psi(g), \lambda^*(g))^\top$. Here, the first two elements represent the intercept and slope terms in the GEV location parameter for the counterfactual world, and the third and fourth entries represent the same for the factual world. The fifth entry denotes the common scale parameter, but on a logarithmic scale. The sixth entry is a specialized transformation of the common shape parameter $\xi(g)$ explained in \eqref{eq:xi2psi}. The seventh entry $\lambda^*(g) = \log[\lambda(g)]$ denotes the strength of dependence of the BHR distribution in \eqref{eq:bvhr} for grid cell $g$, but on a logarithmic scale. 

We next present our proposed framework for quantifying causal effects, which extends the potential outcome paradigm \citep{holland1986statistics} and incorporates marginal return levels. Using expressions of marginal return level in \eqref{eq:rtr_lvl} and model specifications in \eqref{eq:pot}, we represent the return levels in both worlds as 
\begin{equation}
 \nonumber   \begin{aligned}
        Y_t^p(g,0) = \alpha_0(g) + \alpha_1(g)t^*+\frac{\sigma(g)}{\xi(g)}\biggl[\{-\log(1-p)\}^{-\xi(g)}-1\biggl],\\
        Y_t^p(g,1) = \beta_0(g) + \beta_1(g)t^*+\frac{\sigma(g)}{\xi(g)}\biggl[\{-\log(1-p)\}^{-\xi(g)}-1\biggl],    
    \end{aligned}
\end{equation}
where $Y_t^p(g,0)$ is the marginal return level of the grid cell $g$ in the counterfactual world and $Y_t^p(g,1)$ denotes the same in the factual world, $p \in (0,1)$ is the excess level of interest. Now, the causal effect of anthropogenic forcing on high-temperature extremes in the grid $g$ is quantified by the $p$-th return level treatment effect (pRTE), which we defined as
\begin{equation}\label{pRTE}
\nonumber \delta_t(g) = Y_t^p(g,1) - Y_t^p(g,0) = \beta_0(g)-\alpha_0(g) + t^* (\beta_1(g)-\alpha_1(g)).
\end{equation}
Here, since $\delta_t(g)$ depends on $t$, to obtain a single summary of the causal effect of anthropogenic forcing per grid cell, we consider the average over the observational period, i.e., $1850-2014$. Hence, the final causal effect on the grid $g$ is defined as
\begin{equation}\label{pRTE_apprx}
    \delta(g) := \frac{1}{T} \sum_{t=1}^{T} \biggl\{(\beta_0(g)-\alpha_0(g)) + t^* (\beta_1(g)-\alpha_1(g)) \biggl \}.
\end{equation}

\subsection{Assumptions} \label{subsec:assumptions}

We first outline the modeling and notational setup that underpins the application of the potential outcome framework. Here, $Y_t(g)$ is the outcome variable of interest at the grid cell $g$ and year $t$. The presence of anthropogenic forcing resembles an assignment of a treatment variable. Suppose $T_t(g)$ is the treatment variable of interest under the same domain and time setting. Because the treatment assignment is binary, we obtain the outcome variable $Y_t(g)$ under a single treatment. Under two different treatment regimes, i.e., $T_t(g) \in \{0,1\}$, the response variables obtained, $Y_t(g,0)$ and $ Y_t(g,1)$, are generally called potential outcomes. In most cases, the difficulty in estimating the treatment effect arises from the possibility of observing only one of the two potential outcomes per unit, a problem also known as the fundamental problem of causal inference \citep{holland1986statistics}. Therefore, this motivates the use of certain assumptions to identify the causal effect. However, the problem of observing only one potential outcome per unit, which is the grid in our case, is solved by considering numerical climate model output in the presence of anthropogenic forcing as one potential outcome and the output in the absence of anthropogenic forcing as the other one, coupled with the fact that we observe both potential outcomes for each $g \subset \mathcal{D}$ and $t$. We rely on this particular assumption for the identifiability of the proposed treatment effect metric, i.e., pRTE, which is popularly known as the Stable Unit Treatment Value Assumption (SUTVA). Under SUTVA, treatment assignment for one unit does not affect the potential outcomes of other units, implying no interference between units. 

In our attribution structure, using numerical climate model outputs from different experiments to examine the effect of anthropogenic forcing on changes in the intensity of high-temperature extremes, it is plausible to assume that there are no hidden confounders. Because anthropogenic forcing is exogenous, there is no credible factor that both affects it and independently drives the high-temperature extreme. Here, consideration of the spatial independence of annual temperature maxima in both the factual and counterfactual worlds, along with the framing of the problem in a gridded type data setup, makes it convincing to assume that the absence of anthropogenic forcing at one particular grid does not affect the triggering of the occurrence of extreme events in other grids, which in particular implies the satisfaction of the SUTVA assumption. After successive contributions of all these assumptions, the $p$-th return level treatment effect (pRTE), i.e., $\delta(\cdot)$, is estimable.

\subsection{Bayesian hierarchical model} \label{subsec:BHM}

In our proposed hierarchical model, the primary block is the modeling of the observation layer $\mathbf{Y}_t=(\mathbf{Y}_t(g_1),\cdots,\mathbf{Y}_t(g_{250}))^\top,$ where $\mathbf{Y}_t(g_i)=(Y_t(g_i,0), Y_t(g_i,1))^\top$ for $i \in \mathcal{I}=\{1,\cdots,250\}$. Here, $t$ varies across the indexes of the observational period $\mathcal{T}=\{1850,\cdots,2014\}$. We deploy the GEV distribution for marginal modeling of extremes in both worlds and the BHR distribution for joint modeling. The model specifications are detailed in the previous section. The latent layer, which contains the transformed parameters of the BHR distribution, can be represented by $\boldsymbol{\eta} = (\boldsymbol{\eta}(g_1)^\top,\cdots,\boldsymbol{\eta}(g_{250})^\top)^\top$, where $\boldsymbol{\eta}(g_i) = (\alpha_0(g_i),\alpha_1(g_i),\beta_0(g_i), \beta_1(g_i), \sigma^*(g_i),\psi(g_i),\lambda^*(g_i))^\top$. In the context of the areal data setup, a natural option for jointly modeling these latent variables is a multivariate conditional autoregressive (MCAR) model. Here, we prefer the intrinsic CAR (ICAR) model over the usual CAR model due to difficulties with the latter. The difficulty arises from the nature of the estimated autoregression parameter, which often takes values near the limit at which the precision matrix becomes non-positive-definite \citep{banerjee2003hierarchical,cooley2010spatial}. For modeling the mean component in our ICAR model, we select suitable covariates; they include the longitude and latitude of the centroid of the grid cell, the mean elevation of the grid cell, and the mean distance to open sea, following a similar approach adapted by \cite{hrafnkelsson2012spatial} in their work featuring spatial modeling of temperature maxima. The exploratory analysis in Section \ref{sec:EDA} justifies the choice of the covariates. The joint model for the latent layer is represented as
\begin{equation}
    \boldsymbol{\eta} | \boldsymbol{\gamma},\boldsymbol{\Sigma} \sim \mathrm{MVN} (\mathbf{X}\boldsymbol{\gamma},(\mathbf{D-W})^{-1} \otimes \boldsymbol{\Sigma}),
\end{equation}
where $\mathbf{X}$ is the covariate matrix of dimension $1750 \times 35$ in our application, $\boldsymbol{\gamma}$ is the coefficient vector of length $35$, and $\otimes$ denotes the Kronecker product. Here, $\mathbf{X}$ is expressed as
\[
\mathbf{X} =
\left[
\begin{array}{c|c|c|c}
\mathbf{X}_1^\top&\mathbf{X}_2^\top&\cdots&\mathbf{X}_{250}^\top
\end{array}
\right]^\top,
\]
where each block $\mathbf{X}_{i}$ is expressed as $\mathbf{X}_i = \mathbf{I}_{7} \otimes \widetilde{\mathbf{X}}_i^\top$ and $\widetilde{\mathbf{X}}_i =(1,\widetilde{X}_i^{(1)}, \cdots, \widetilde{X}_i^{(4)})^\top$. Here, $\widetilde{\mathbf{X}}_i$ contains all the necessary details about the covariates for grid $g_i$. More specifically, $\widetilde{X}_i^{(1)}$ is latitude corresponding to the centroid of grid $g_i$, $\widetilde{X}_i^{(2)}$ is longitude corresponding to the centroid of grid $g_i$, $\widetilde{X}_i^{(3)}$ is the mean elevation corresponding to the grid $g_i$, and $\widetilde{X}_i^{(4)}$ is the proximate sea distance corresponding to the grid $g_i$; with $ \hspace{1.5mm} g_i \subset \mathcal{D}=g_1 \cup\cdots\cup g_{250}$. Further, $\boldsymbol{\Sigma}$ represents the $7 \times 7$-dimensional cross-component covariance matrix within a particular grid cell, and $\mathbf{W}$ is the symmetric proximity matrix, whose $(i,j)^\text{th}$ element is given by
\begin{equation}\nonumber
    w_{ij}=
    \begin{cases}
        1,& \text{if} \quad g_i \sim g_j \\
        0,&  \text{otherwise},
    \end{cases}
\end{equation}
where $g_i \sim g_j$ represents the adjacency of the grid cells $g_i$ and $g_j$ and $w_{ii} = 0, \hspace{1.5mm} \forall g_i \subset \mathcal{D}$. Here, $\mathbf{D}$ is the diagonal matrix whose diagonal elements represent the total number of adjacent regions to a particular region, i.e., $ d_{ii} = \sum _{j \sim i} w_{ij}, \hspace{1.5mm} \forall g_i,g_j \subset \mathcal{D}$. Finally, we choose weakly-informative conjugate priors for the hyperparameters that the latent variables depend on. We consider a multivariate normal prior for the regression coefficient vector $\boldsymbol{\gamma}$, given by $\boldsymbol{\gamma} \sim \mathrm{MVN}(\mathbf{0},\sigma^{2}_{\boldsymbol{\gamma}}\mathbf{I}_{35})$, and we select an inverse-Wishart prior for the cross-covariance matrix $\boldsymbol{\Sigma}$, given by $\boldsymbol{\Sigma} \sim \mathrm{IW}(\nu,\boldsymbol{\Psi})$, where $\nu$ denotes the degrees of freedom and $\boldsymbol{\Psi}$ is the scale matrix of dimension $7 \times 7$. In our application, we fix $\sigma^{2}_{\boldsymbol{\gamma}} = 100^2$, $\nu=0.1$, and $\boldsymbol{\Psi}=0.1 \mathbf{I}_7$, which ensure the weakly-informative nature of the hyperpriors. After including all the essential modeling details about the data layer (except the shape parameter regularization in the data layer as described in Section \ref{subsec:data_layer}) and also the details about the assigned priors for each layer, the proposed hierarchical Bayesian model is
\begin{equation}\label{eq:overall}
    \begin{split}
    \text{Data layer: } \quad
        & (Y_t(g,0),Y_t(g,1))^\top |\bm{\eta}(g) \overset{\mathrm{ind}}{\sim} \text{BHR} (\bm{\eta}(g)),\\[2mm]
        & \text{where} \hspace{1.5mm} \bm{\eta}(g) = (\alpha_0(g), \alpha_1(g), \beta_0(g), \beta_1(g), \sigma^*(g), \psi(g), \lambda^*(g))^\top,\\[2mm]
        & \text{with} \hspace{1.5mm} g \subset g_1 \cup\cdots\cup g_{250} \subset \mathcal{D}, \hspace{1.5mm} t \in \mathcal{J}=\{1,\cdots,165\}, \\
        \\
    \text{Latent layer: } \quad 
        & \boldsymbol{\eta} | \boldsymbol{\gamma},\boldsymbol{\Sigma} \sim \mathrm{MVN} (\mathbf{X}\boldsymbol{\gamma},(\mathbf{D-W})^{-1} \otimes \boldsymbol{\Sigma}), \\
        \\
    \text{Prior layer: } \quad
        & \boldsymbol{\gamma} \sim \mathrm{MVN}(\mathbf{0},\sigma^{2}_{\boldsymbol{\gamma}}\mathbf{I}_p),~ \boldsymbol{\Sigma} \sim \mathrm{IW}(\nu,\boldsymbol{\Psi}).
    \end{split}
\end{equation}

The notations used in \eqref{eq:overall} are as defined in Sections \ref{subsec:causal_inference} and \ref{subsec:BHM}.

\section{Bayesian inference and hotspot identification}\label{sec:inference_hotspot}

\subsection{Approximate Bayesian inference method}\label{subsec:ABI}

Here, we adopt a similar approach to \cite{johannesson2022approximate} for Bayesian inference, where we approximate the likelihood function with a Gaussian density and combine it with the latent layer to obtain a Gaussian-Gaussian pseudo-model. Building on this framework and using the proposed model in \eqref{eq:overall} as well as the posterior density expression in \eqref{eq:posterior}, we now show that the posterior density can be written as
\begin{equation}\label{post}
    \pi(\boldsymbol{\eta,\gamma,\Sigma} | \mathbf{y}) \varpropto \pi(\boldsymbol{\Sigma})\pi(\boldsymbol{\gamma})\pi(\boldsymbol{\eta|\gamma,\Sigma)}\pi(\mathbf{y}|\boldsymbol{\eta})
     \varpropto \pi(\boldsymbol{\Sigma})\pi(\boldsymbol{\gamma})\pi(\boldsymbol{\eta|\gamma,\Sigma)}L(\boldsymbol{\eta}|\mathbf{y}),
\end{equation}
where $\mathbf{y}$ denotes the observed data vector combining all grid cells and years, and $L$ denotes the generalized likelihood function, 
with $L(\boldsymbol{\eta}|\mathbf{y})$ is expressed as 
\begin{equation}\label{eq:Likelihood}
    L(\boldsymbol{\eta}|\mathbf{y}) = \prod_{t=1}^{165} \prod_{i=1}^{250}L((y_t(g_i,0),y_t(g_i,1))'|\boldsymbol{\eta}(g_i)) \pi(\psi(g_i)).
\end{equation}
Here, $\pi(\psi(g_i))$ is as defined in Section \ref{subsec:data_layer} and presented in the right panel of Figure \ref{fig:prior.psi}, and used in regularization in the first stage. To avoid confusion, similar to \cite{johannesson2022approximate}, here we use $\pi(\psi(g_i))$ for regularization in the first stage and then consider an appropriate (ICAR) multivariate Gaussian prior structure for $\psi(g_i)$s in the latent layer as well. One may view the product of these two terms, i.e., $\pi(\psi(g_i))$ for $i=1,\ldots, 250$ and the multivariate Gaussian density as the final prior for $\psi(g_i)$s. Now, the generalized likelihood function is approximated by a Gaussian density having mean $\widehat{\boldsymbol{\eta}}$ and the covariance matrix $\boldsymbol{\Sigma_{\widehat{\eta}}}$, where $\widehat{\boldsymbol{\eta}} = \operatorname*{arg\,max}_{\boldsymbol{\eta}} \log L(\boldsymbol{\eta}|\mathbf{y})$ and $\boldsymbol{\Sigma_{\widehat{\eta}}}$ is the inverse of negative Hessian matrix of $\log L$, evaluated at $\widehat{\boldsymbol{\eta}}$. Then, the approximated likelihood function $\widehat{L}(\widehat{\boldsymbol{\eta}}|\mathbf{y})$ can be written as $\widehat{L}(\widehat{\boldsymbol{\eta}}|\mathbf{y}) = \mathcal{N}(\boldsymbol{\eta} | \widehat{\boldsymbol{\eta}},\boldsymbol{\Sigma_{\widehat{\eta}}})$. Here, $\widehat{\bm{\eta}}$ is a 1750-length vector and $\boldsymbol{\Sigma_{\widehat{\eta}}}$ is a $1750\times1750$-dimensional matrix. However, due to assuming independence across grid cells in the data layer, $\boldsymbol{\Sigma_{\widehat{\eta}}}$ is highly sparse and block diagonal in nature, with 250 blocks each of dimension $7\times7$, corresponding to seven parameters in $\bm{\eta}(g_i)$. Hence, the approximate posterior density can be written as
\begin{equation}\label{apprx-post}
      \pi(\boldsymbol{\eta,\gamma,\Sigma} | \widehat{\boldsymbol{\eta}})
      \varpropto \pi(\boldsymbol{\gamma})\pi(\boldsymbol{\Sigma})\pi(\boldsymbol{\eta|\gamma,\Sigma)}\pi(\widehat{\boldsymbol{\eta}}|\boldsymbol{\eta})
      \varpropto \pi(\boldsymbol{\Sigma})\pi(\boldsymbol{\gamma})\pi(\boldsymbol{\eta|\nu,\Sigma)}\mathcal{N}(\boldsymbol{\eta} | \widehat{\boldsymbol{\eta}},\boldsymbol{\Sigma_{\widehat{\eta}}}).
\end{equation}
This approximation by $\pi(\boldsymbol{\eta,\gamma,\Sigma} | \widehat{\boldsymbol{\eta}})$ to our exact posterior density $\pi(\boldsymbol{\eta,\gamma,\Sigma} | \mathbf{y})$ leverages conjugacy, making posterior inference easier and faster than with the exact model. Here, we apply the Bayesian inference scheme proposed by \cite{hrafnkelsson2021max}, i.e., Max-and-Smooth, where the process of approximating the generalized likelihood function with a Gaussian density with mean $\widehat{\boldsymbol{\eta}}$ and the covariance matrix $\boldsymbol{\Sigma_{\widehat{\eta}}}$ refers to the Max-step and inference of the latent variables and hyperparameters, and smoothening the latent surface using the Gaussian-Gaussian pseudo model refers to the Smooth-step. The advantages of the Max-and-Smooth approach are that the computational cost of implementing a Markov chain Monte Carlo (MCMC) algorithm remains roughly constant even as the number of independent replications (here, years) of the observed data increases. Besides, the sparse precision matrix of the Gaussian prior density is beneficial for computational efficiency. 


\subsection{MCMC implementation} \label{subsec:mcmc}
Obtaining the Gaussian-Gaussian pseudo-model as the end product of the extended Max-and-Smooth approach, along with reasonable prior choices for hyperparameters, enables us to implement Gibbs sampling to draw posterior inference about the latent variables and hyperparameters. While updating $\boldsymbol{\eta}$ of length 1750 using Gibbs sampling, the sparse precision matrix of Kronecker-form and the sparse structure of $\mathbf{X}$ enable using the \texttt{R} package \texttt{spam} \citep{furrer2010spam}, and its appropriate tools drastically increase the speed of the Gibbs sampling steps. Similarly, while updating $\boldsymbol{\gamma}$, we rely on the \texttt{spam} package for boosting the speed up due to the involvement of $\mathbf{X}$ in it. The conjugacy feature, enabled by the inverse-Wishart prior for $\boldsymbol{\Sigma}$, facilitates Gibbs sampling updates for it. Overall, our proposed model provides closed-form expressions for the full conditional posterior distributions of all model parameters and hyperparameters, thereby avoiding any Metropolis-Hastings steps. The full conditional posteriors (FCPs) for updating $\boldsymbol{\eta}$ of length 1750, $\boldsymbol{\gamma}$ of length 35, and $\boldsymbol{\Sigma}$ of dimension $7\times 7$ in our Gibbs sampling scheme are as follows:
\begin{itemize}
    \item Simulate $\boldsymbol{\eta}$ from the FCP of $\bm{\eta}$, given by $\boldsymbol{\eta}|\mathbf{y},\mathbf{X},\boldsymbol{\gamma},\boldsymbol{\Sigma} \sim \mathrm{MVN} (\boldsymbol{\mu}_{\boldsymbol{\eta}}^{\text{post}},\boldsymbol{\Sigma}_{\boldsymbol{\eta}}^{\text{post}})$,  where\\$\boldsymbol{\mu}_{\boldsymbol{\eta}}^{\text{post}}=\boldsymbol{\Sigma}_{\boldsymbol{\eta}}^{\text{post}} (\boldsymbol{\Sigma}_{\widehat{\boldsymbol{\eta}}}^{-1}\widehat{\boldsymbol{\eta}}+[(\mathbf{D}-\mathbf{W})\otimes \boldsymbol{\Sigma}^{-1}] \mathbf{X}\boldsymbol{\gamma})$ and  
    $\boldsymbol{\Sigma}_{\boldsymbol{\eta}}^{\text{post}} = (\boldsymbol{\Sigma}_{\widehat{\boldsymbol{\eta}}}^{-1} +(\mathbf{D}-\mathbf{W}) \otimes \boldsymbol{\Sigma}^{-1})^{-1}$,
    \item Simulate $\boldsymbol{\gamma}$ from its FCP, $\boldsymbol{\gamma}|\boldsymbol{\eta},\mathbf{X},\boldsymbol{\Sigma}  \sim \mathrm{MVN} (\boldsymbol{\mu}_{\boldsymbol{\gamma}}^{\text{post}},\boldsymbol{\Sigma}_{\boldsymbol{\gamma}}^{\text{post}})$, where \\
    $\boldsymbol{\mu}_{\boldsymbol{\gamma}}^{\text{post}} = \boldsymbol{\Sigma}_{\boldsymbol{\gamma}}^{\text{post}} (\mathbf{X}^\top [(\mathbf{D}-\mathbf{W})\otimes \boldsymbol{\Sigma}^{-1}]\boldsymbol{\eta})$ and $\boldsymbol{\Sigma}_{\boldsymbol{\gamma}}^{\text{post}} = 
    (\mathbf{X}^\top[(\mathbf{D}-\mathbf{W})\otimes\boldsymbol{\Sigma}^{-1}]\mathbf{X}+\sigma^{-2}_{\boldsymbol{\gamma}}\mathbf{I}_{35})^{-1}$,
    \item Simulate $\boldsymbol{\Sigma}$ from its FCP, $\boldsymbol{\Sigma}|\boldsymbol{\eta},\mathbf{X},\boldsymbol{\gamma} \sim \text{IW} (\nu^{\text{post}}, \boldsymbol{\Psi}^{\text{post}})$, where \\
    $\nu^{\text{post}}=\nu+|\mathcal{I}|$ and $\boldsymbol{\Psi}^{\text{post}}= \operatorname{vec}^{-1}(\boldsymbol{\eta}-\mathbf{X}\boldsymbol{\gamma})(\mathbf{D}-\mathbf{W}) (\operatorname{vec}^{-1}(\boldsymbol{\eta}-\mathbf{X}\boldsymbol{\gamma}))^\top + \boldsymbol{\Psi}$.    
\end{itemize}
Here, $\operatorname{vec}^{-1}$ converts the vector $\boldsymbol{\eta}-\mathbf{X}\boldsymbol{\gamma}$ of length 1750 into a $7\times 250$ matrix, with the first seven entries of the vector forming the first column of the matrix, the next seven entries forming the second column of the matrix, and so on. In our data application, we implement the MCMC algorithm in \texttt{R} (\hyperlink{http://www.r-project.org/}{http://www.r-project.org/}). We generate $60,000$ posterior samples and discard the first $10,000$ samples as burn-in. After choosing the thinning equal to $5$, that is, keeping one out of every five consecutive samples of the Markov chain, we are left with $10,000$ samples. These $10,000$ samples are used for drawing posterior inference. The convergence and mixing of the chains are well-examined through their trace plots, Geweke statistics, and Gelman-Rubin diagnostics.

\subsection{Hotspot estimation} \label{subsec:hotspot}

In addition to estimating the effect of human intervention on high temperature extremes, our goal is to identify the regions at risk, that is, to construct a credible region that contains the true exceedance set with a specified level of probability. Here, the true exceedance set refers to the collection of grid cells where the estimated treatment effect exceeds the chosen threshold. We adapt the proposal based on the Bayesian setting of \cite{french2016credible} for outer credible region construction, which is the natural extension of the frequentist setting proposed by \cite{french2013spatio}. The primary reason for deploying the sampling-based methodology of \cite{french2016credible} is that it is easy to implement, as it relies on samples from the posterior predictive distribution generated by MCMC methods.

Let $ s_{g_i}$ be the centroid, which is also known as the representative location, corresponding to the grid $g_i$, where $i \in \mathcal{I}=\{1,\cdots,250\}$. For an user-defined threshold $u$, the true exceedance region is defined as $E_{u^+}= \{g: \delta(s_g) > u \}$, where $\delta(\cdot)$ is the causal effect and $u$ is the chosen threshold. Our goal is to construct the outer-credible region, i.e., $C_{u^+}^{O} (\alpha)$ such that $\mathrm{P}(E_{u^+} \subset C_{u^+}^{O} (\alpha)) = 1-\alpha$. The bridge between obtaining samples from the posterior predictive distribution and constructing $C_{u^+}^{O}$ is built on a multiple-hypothesis testing procedure. However, the outer credible region $C_{u^+}^{O}$ is not unique and depends on the discretization of the domain of interest. For a gridded dataset like ours, the discretization is pre-specified. A major drawback of testing $H_0(s_g):\delta(s_g) \geq 0$ versus $H_1(s_g):\delta(s_g) < 0$ individually in each grid cell based on some decision rule $\phi(s_g)$, which depends upon the test statistics $T(s_g)$ with predefined confidence level $1-\alpha$, is the failure in accurate representation of joint threshold exceedances. To address this issue, we perform multiple hypothesis testing by selecting a suitable critical value that controls the family-wise error rate at level $\alpha$. In our analysis, to perform the multiple hypothesis testing, we choose the grid cell-wise test statistics
\begin{equation}
    T(s_g)=\frac{\mathbb{E}(\delta(s_g)|\mathbf{y})-u}{\sqrt{\mathbb{E}((\delta(s_g)-u)^2|\mathbf{y})}}.
\end{equation}
The detailed description of the algorithm for estimating the region containing the true joint outer exceedance region, i.e., $C_{u^+}^{O} (\alpha)$, is given in Algorithm \ref{alg:alg1}.

\begin{algorithm}[t] 
    \caption{Hotspot estimation algorithm \citep{french2016credible}}
    \label{alg:alg1}
    \begin{itemize}
        \item Fix the threshold level $u$ and the family-wise error rate $\alpha$.
        \item Obtain $B$ MCMC samples from the approximate posterior (using Max-and-Smooth) corresponding to the latent variables and hyperparameters.
         \item for each posterior samples $b \in \{1,\cdots,B\}$:
         \begin{itemize}
             \item Retrieve posterior samples  $\alpha_0^{(b)}(s_g),\beta_0^{(b)}(s_g),\alpha_1^{(b)}(s_g),\beta_1^{(b)}(s_g)$, where $g \subset \mathcal{D}$.
             \item Calculate $\delta^{(b)}(s_g)$ using \eqref{pRTE_apprx}.
         \end{itemize}
        \item Calculate the test statistics corresponding to each grid, i.e., $T(s_{g_{1}}),\cdots,T(s_{g_{250}})$.
        \item Using $\widetilde{\delta}^{(b)}$, where $\widetilde{\delta}^{(b)} = \{ \delta^{(b)}(s_{g_{1}}),\cdots,\delta^{(b)}(s_{g_{250}})\}$ and $b \in \{1,\cdots,B\}$:
        \begin{itemize}
            \item Identify the exceedance set above the threshold $u$, i.e., $E_{u^+}^{(b)}=\{g_i:\delta^{(b)}(s_{g_{i}}) \geq u, ~i\in\{1,\cdots,250\} \}.$
            \item Determine $\mathcal{M}^{(b)}$, where $\mathcal{M}^{(b)}=\text{min}\{T(s_g);g \in E_{u^+}^{(b)}\}$. Set $\mathcal{M}^{(b)} = 0$, if $E_{u^+}^{(b)}$ is empty.
        \end{itemize}
        \item Estimate the critical value $c(\alpha)$ using the $\alpha$-quantile of $\{ \mathcal{M}^{(1)},\cdots,\mathcal{M}^{(B)}\}.$
        \item Obtain $C_{u^+}^{O}$, where $C_{u^+}^{O} (\alpha) =\{g_i:T(s_{g_{i}})\geq \widehat{c}(\alpha)\}$.
    \end{itemize}
\end{algorithm}
\section{Data application}\label{sec:data_application}

We fit the proposed Bayesian hierarchical model in Section \ref{subsec:BHM} to the climate model output described in Section \ref{sec:EDA}. Utilizing the approximate Bayesian inference method in Section \ref{subsec:ABI}, we obtain $10,000$ posterior samples for each latent variable and hyperparameter. Although trace plots of Markov chains for different parameters are widely used to assess convergence and mixing, practitioners often use multiple diagnostic tools to assess convergence and shed light on the quality of posterior samples. Given that the number of latent variables and hyperparameters is high in our model, we focus on the diagnostics for the hyperparameter vector $\bm{\gamma}$. Gelman-Rubin diagnostics, denoted by $R$-value, assess the quality of posterior samples and indicate agreement between posterior means across multiple chains; we obtain the estimated $R$-value, i.e., $\widehat{R}$ and its upper confidence limit for each MCMC chain. Generally, $\widehat{R}=1$ indicates convergence of the chain, and a value of $\widehat{R}>1.1$ indicates questionable convergence \citep{reich2019bayesian}. In our case, the $\widehat{R}$ values are equal to 1, and also the upper confidence limits are equal to 1 for all the hyperparameters in $\boldsymbol{\gamma}$; they signify the convergence of MCMC chains. Further, Geweke's diagnostic is also intended to assess the convergence of the Markov chains for the hyperparameters in $\boldsymbol{\gamma}$. Geweke's statistic is essentially a $z$-score that signifies the difference between the sample means of two subsequences of the Markov chain, divided by the estimated standard error. Hence, an absolute value of the score below 1.96 indicates convergence of the Markov chain. The calculated Geweke's statistics for all the hyperparameters stored in $\boldsymbol{\gamma}$ are reported in the top block of Table~\ref{Geweke_ESS_combined}. The score values lie within the 0.025$\mathrm{th}$ and 0.975$\mathrm{th}$ standard normal quantiles for all considered hyperparameters, indicating that the MCMC chains have converged after the burn-in period. Further, the Effective sample size (ESS) quantifies the number of independent samples that contain the same information as the total number of correlated samples in the Markov chain. These pseudo-independent samples are used to improve the uncertainty assessment of the posterior mean calculated from posterior samples. In the bottom block of Table \ref{Geweke_ESS_combined}, the ESS of all 35 hyperparameters in $\boldsymbol{\gamma}$ are reported. Overall, the ESS values are on the higher side for all the model hyperparameters except one coefficient related to $\lambda^*$. All the desired metrics obtained from diagnostic tools, such as Geweke's statistic, Gelman-Rubin statistics, and reported higher ESS values, collectively imply the good quality of the posterior samples used for drawing posterior inference.

\begin{table}[h]
\centering
\caption{Geweke $z$-scores (top) and effective sample sizes (bottom) for all hyperparameters in $\boldsymbol{\gamma}=(\boldsymbol{\gamma}_{\alpha_{0}},\boldsymbol{\gamma}_{\alpha_{1}},\boldsymbol{\gamma}_{\beta_{0}},\boldsymbol{\gamma}_{\beta_{1}},\boldsymbol{\gamma}_{\sigma^{*}},\boldsymbol{\gamma}_{\psi},\boldsymbol{\gamma}_{\lambda^*})'$, where $\boldsymbol{\gamma}_{\textit{j}} = (\gamma^{(1)}_{\textit{j}},\gamma^{(2)}_{\textit{j}},\gamma^{(3)}_{\textit{j}}, \gamma^{(4)}_{\textit{j}}, \gamma^{(5)}_{\textit{j}})', \forall \textit{j} \in \{ \alpha_{0},\alpha_1,\beta_0,\beta_1,\sigma^*,\psi,\lambda^*\} $. Here, each element of the table corresponds to the Geweke statistic or the effective sample size for a specific component of the vector of covariate coefficients associated with latent variables.}
\begin{tabular}{|l|ccccccc|}
\multicolumn{1}{c}{}
 &
$\boldsymbol{\gamma}_{\alpha_0}$ &
$\boldsymbol{\gamma}_{\beta_0}$ &
$\boldsymbol{\gamma}_{\alpha_1}$ &
$\boldsymbol{\gamma}_{\beta_1}$ &
$\boldsymbol{\gamma}_{\sigma^*}$ &
$\boldsymbol{\gamma}_{\psi}$ &
\multicolumn{1}{c}{$\boldsymbol{\gamma}_{\lambda^*}$} \\
    \hline
\multicolumn{8}{|c|}{Geweke $z$-scores} \\
\hline
$\gamma^{(1)}$ & -0.46 & 0.45 & 0.13 & -0.42 & -0.97 & -0.45 & 0.19\\
$\gamma^{(2)}$ & 0.58 & -0.60 & 0.71 & 1.81 & -0.66 & 1.32 & 0.30 \\
$\gamma^{(3)}$ & -0.41 & -0.08 & -0.37 & -0.35 & 1.97 & 0.07 & 0.17 \\
$\gamma^{(4)}$ & 0.81 & 0.55 & 0.88 & -0.11 & -0.47 & 0.59 & 1.52 \\
$\gamma^{(5)}$ & 0.30 & 0.05 & 0.34 & 0.94 & -1.41 & 0.22 & -0.66 \\
\hline
\multicolumn{8}{|c|}{Effective Sample Size (ESS)} \\
\hline
$\gamma^{(1)}$ & 12000 & 12000 & 12000 & 12000 & 12000 & 12000 & 12000 \\
$\gamma^{(2)}$ & 12000 & 12000 & 12000 & 12000 & 10576 & 12000 & 3863 \\
$\gamma^{(3)}$ & 12000 & 11389 & 12000 & 12000 & 8365 & 9560  & 2689 \\
$\gamma^{(4)}$ & 12000 & 5712 & 12000 & 12000 & 2326 & 2764 & 702 \\
$\gamma^{(5)}$ & 12000 & 10696 & 12000 & 12000 & 5770 & 7978 & 1952 \\
\hline
\end{tabular}
\label{Geweke_ESS_combined}
\end{table}

Before moving to discussing inferences about the grid cell-level parameters, we summarize the posterior means and standard deviations of all model hyperparameters in $\boldsymbol{\gamma}$ in Table \ref{post_mean_latent}. The intercept terms have very large posterior standard deviations relative to their means, indicating substantial uncertainty and suggesting weak identifiability or limited information in the data regarding baseline effects. In contrast, covariates such as longitude and latitude demonstrate pronounced and consistent effects across most latent variables, with relatively smaller standard deviations, underscoring their strong spatial influence. Latitude, in particular, is associated with large magnitude coefficients and moderate uncertainty, reflecting a dominant north–south gradient. Mean elevation (ME) exhibits stable and precisely estimated effects, as indicated by very small standard deviations, which demonstrates a robust and consistent contribution across components. Open sea distance (OS) also displays clear effects, though with slightly higher variability than ME, suggesting a moderate influence with some uncertainty. Collectively, these results indicate that spatial covariates, especially latitude and longitude, are key components of the latent structure, while elevation provides a steady contribution, and intercept terms remain weakly identified.

\begin{table}[h]
    \centering
    \caption{Posterior means and standard deviations (within brackets) of all the hyperparameters stored in $\bm{\gamma}$. Here, different columns represent the posterior means and standard deviations of the covariate coefficients for different latent variables. Each row represents the posterior means and standard deviations of the coefficients for a specific covariate. Here, Int denotes the intercept, Lon denotes the Longitude, Lat denotes the latitude, ME denotes the mean elevation, and OS denotes the open sea distance. }
    \setlength{\tabcolsep}{0.3pt}
    \begin{tabular}{|l|c|c|c|c|c|c|c|}
    \multicolumn{1}{c}{}&
    $\alpha_0$ &  
    $\alpha_1$ &
    $\beta_0$ &
    $\beta_1$ &
    $\sigma^*$ &
    $\psi$  &
    \multicolumn{1}{c}{$\lambda^*$} \\
    \hline   
    Int   & -1.53(99.07)  & 19.41(101.13) & -4.07(100.15)  & -4.83(99.09)  & 2.43 (100.19) & -8.67(99.41)  & -9.41(100.63)\\
    Lon   &-455.0(50.27)  &14.29(2.57) &-438.87(49.56) &-8.63(4.41)  &-71.60(3.88)  &30.79(2.27) & 44.39(6.52)\\
    Lat  &-3654.42(56.45)  & 82.74(3.47)  &-3829.63(55.82)  &72.23(5.05)  &140.55(5.46)  &80.31(3.01)  &85.87(12.01)\\
    ME   &-51.40(0.21)  &0.17(0.02) &-50.39(0.20)  &0.17(0.02)  &-1.75(0.04) &0.13(0.01)  &0.85(0.09)\\
    OS   &30.66(0.66)  &0.51(0.04) &28.28(0.65)  &-2.15(0.06)  &1.83(0.07)  &-0.94(0.04)  &-1.65(0.14)\\
    \hline
    \multicolumn{8}{l}{\footnotesize Here, expect the elements of the first row, elements of all the remaining rows are multiplied by $10^{3}$.}
    \end{tabular}
    \label{post_mean_latent}
\end{table}

Our primary goal is to estimate the effect of human intervention on high temperature extremes using the causal metric defined in \eqref{pRTE}. Hence, we utilize the posterior samples of $\alpha_0,\alpha_1,\beta_0,~\text{and}~\beta_1$ to estimate the causal effect, given their role in the final expression of the causal metric. We calculate posterior estimates of the causal effect for each grid that jointly constitute the domain of interest, i.e., the mainland United States. 
\begin{figure}[t]
  \centering 
  \includegraphics[width=\textwidth]{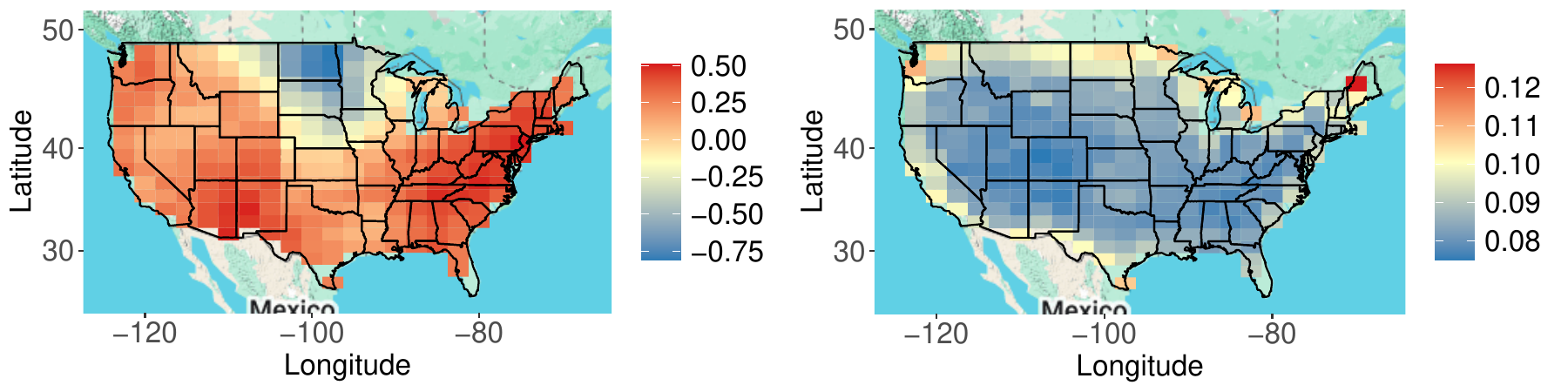}
  \caption{Grid cell-wise posterior means (left) and posterior standard deviations (right) of the causal effect. }
  \label{causal_effect_plot}
\end{figure}
The spatial map of the grid cell-wise posterior means of the causal effect is displayed in the left panel of Figure \ref{causal_effect_plot}. Alongside that, the spatial map of the grid cell-wise posterior standard deviations calculated using posterior samples of causal effects is shown in the right panel of Figure \ref{causal_effect_plot}. From this figure, it is evident that in most regions of the mainland United States, the effect of anthropogenic forcing on high-temperature extremes is positive. In the northeast region and some parts of the southern region, including the South Atlantic and the East South Central division, the causal effect estimate is higher than in all other regions, with the estimated mean causal effect ranging from $0.20$ to $0.50$. In the West region, the West South Central and East North Central divisions, the mean causal effect lies between $0$ and $0.3$, while in a fraction of the subregions it is approximately $0.5$. Nevertheless, in the West North Central division of the Midwest region, which includes North Dakota, Minnesota, South Dakota, Nebraska, some parts of Iowa, and Montana, the posterior mean of the causal effect is negative. While the positive effect implies a significant contribution of anthropogenic forcing to the frequent occurrence of high-temperature extremes, the negative effect implies no such contribution. According to the CENSUS report \citep{hartley2021preliminary}, the major parts of the midwestern states where the estimated causal effect is negative are sparsely populated due to several reasons, including harsh climate conditions, and the agricultural and grazing land covers a substantial portion of the land in the Midwestern region containing North Dakota, South Dakota, Nebraska, and Iowa \citep{pryor2013midwestern}. Natural forcing, such as atmospheric circulation patterns and solar radiation, has a greater impact on the occurrence of high-temperature extremes \citep{andresen2012historical} in these regions than human-induced forcing, such as greenhouse gas emissions. However, the northeastern region, with major metropolitan cities, and the states in the South Atlantic division are densely populated \citep{fonseca2000changing}, which implies land-use change and urbanization and leads to anthropogenic forcing acting as a key contributor to the warmer climate in those areas.
\begin{figure}[t]
  \centering 
  \includegraphics[width=\textwidth]{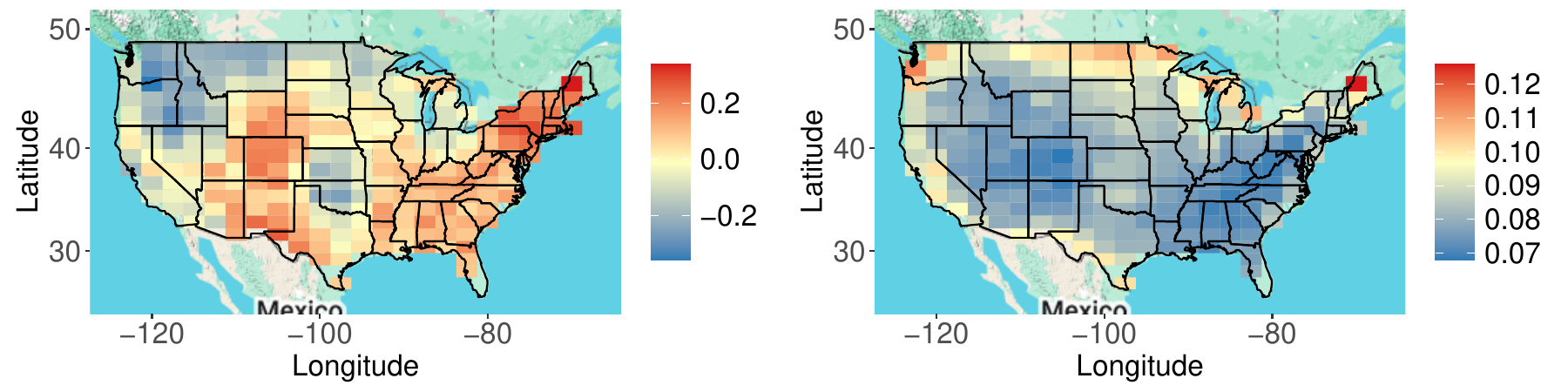}
  \caption{Grid cell-wise posterior means (left) and posterior standard deviations (right) of the causal effect during the pre-industrial period (1850--1900).}
  \label{causal_effect_plot_pi}
\end{figure}

Generally, the period $1850-1900$ is considered the pre-industrial era and is used as a baseline for climate change assessment studies. Figure \ref{causal_effect_plot_pi} displays the spatial maps of the posterior mean causal effect across the mainland United States during this pre-industrial period. The estimated causal effect is negative in most regions of the mainland United States. The comparison between the spatial maps of the estimated causal effect over the whole time period and only the pre-industrial period shows that anthropogenic forcing is a significant contributor to the recurrence of high-temperature extremes.
\begin{figure}[t] 
  \centering
  \includegraphics[width=\textwidth]{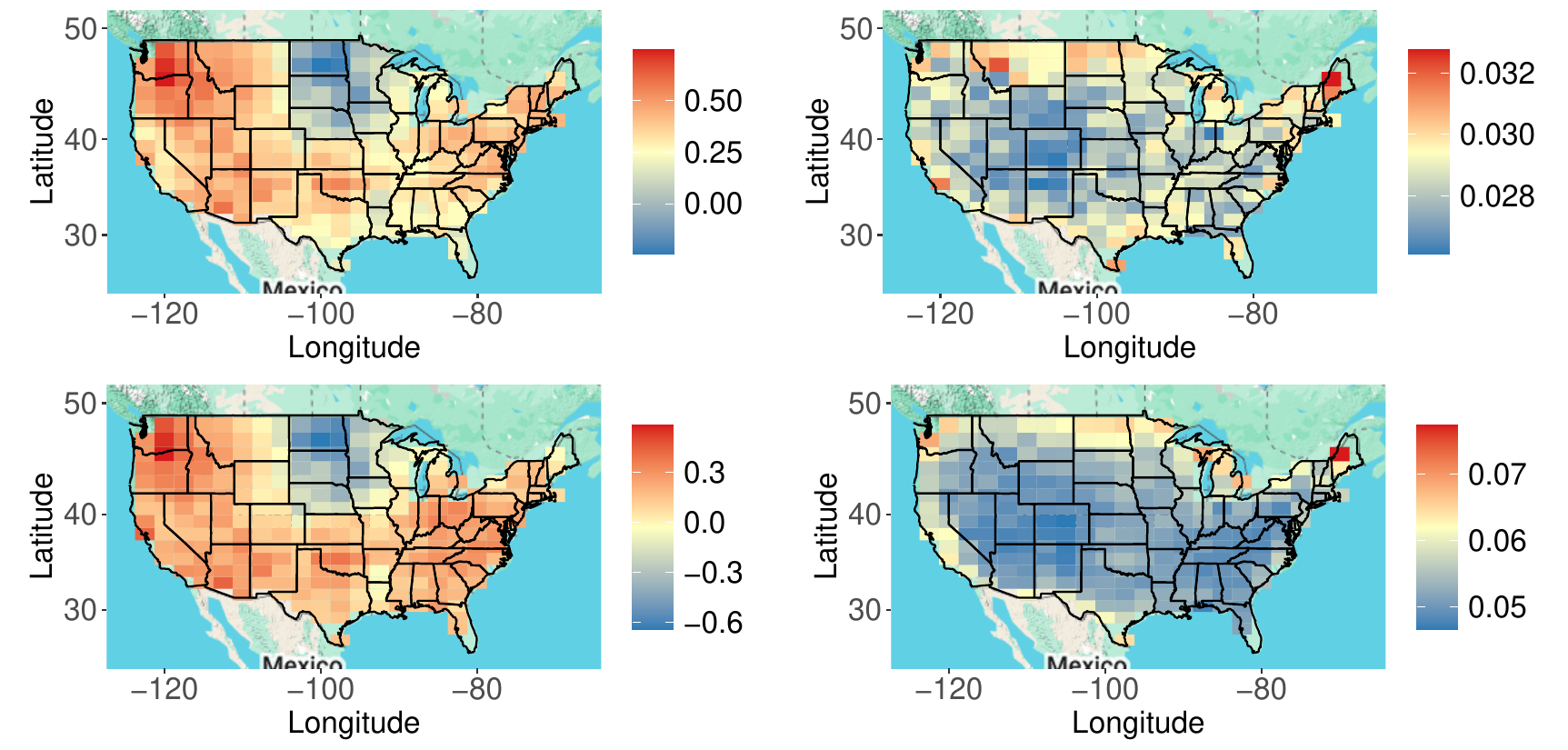}
  \caption{Grid cell-wise posterior means (top-left) and posterior standard deviations (top-right) of the trend (the change in the intensity of high-temperature extremes per unit change in year) for the factual world, and the pixel-wise posterior means (bottom-left) and posterior standard deviations (bottom-right) of the trend difference between the factual and counterfactual worlds.}
  \label{trend_USA_plot}
\end{figure}

The extensive information on the trend of high-temperature extremes helps in designing more effective adaptation and mitigation policies to address them to some extent. Here, the trend is defined as the change in the intensity of high-temperature extremes per unit change in year. We turn to $\alpha_1$ and $\beta_1$ to understand the trend behavior in both counterfactual and factual worlds. Hence, the spatial map of the posterior means of $\beta_1$ is shown in the left panel of Figure \ref{trend_USA_plot}, which presents the trend in the factual world. In addition to this, the spatial map of corresponding standard deviations is also displayed in the right panel of Figure \ref{trend_USA_plot}. In the factual world, the Midwestern region exhibits a negative trend in high-temperature extremes over the year, while all other regions show a positive trend in their intensity due to rapid urbanization, industrial exposure, and land-use changes. Stakeholders and climate policymakers may be highly interested in the observed trend in the factual world. However, the changes in trends of both counterfactual and factual worlds are also shown in Figure \ref{trend_USA_plot}. From the perspective of highlighting the significant contribution of anthropogenic forcing to the intensification of temperature extremes across most regions of the mainland United States, excluding the Midwest, the trend-difference spatial map is considered important. Alongside that, the spatial map of the corresponding standard deviations is presented in Figure \ref{trend_USA_plot}. 
\begin{figure}[t]
  \centering
  \includegraphics[width = \textwidth, height = 4.5cm]{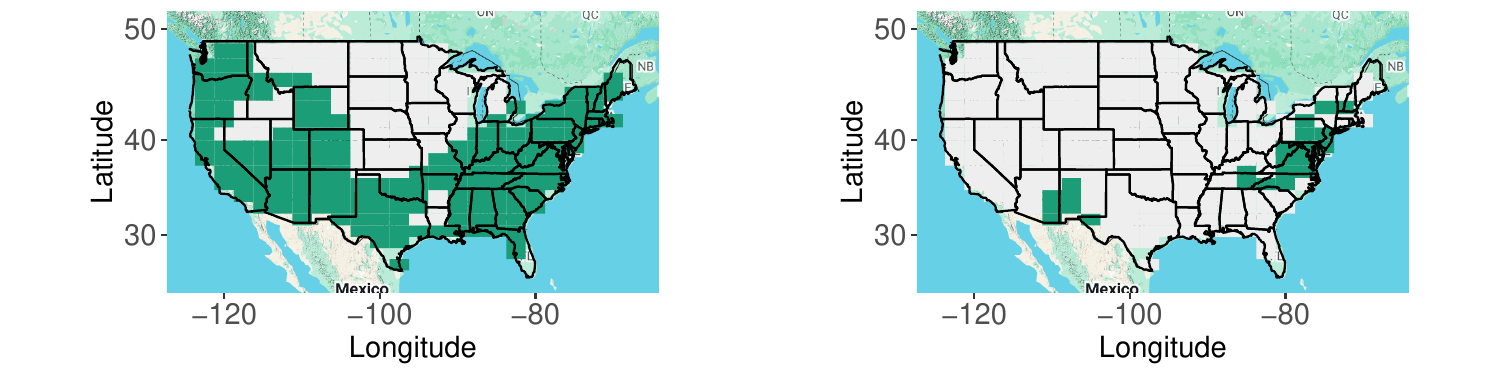}
  \caption{Estimated $95\%$ credible regions $C_{u^{+}}^{O}$ corresponding to the joint exceedance levels, $u = 0.35$ (left) and $u=0.65$ (right).}
  \label{hotspot_USA_plot}
\end{figure}

Considering the importance of identifying regions at risk, we estimate the outer credible region using the hotspot region estimation technique given in Section \ref{subsec:hotspot}. We construct $95\%$ credible regions $C_{u^{+}}^{O}$ corresponding to high threshold levels, $u =0.35$ and $u=0.65$. Figure \ref{hotspot_USA_plot} shows the hotspot regions over the mainland United States. For $u=0.35$, the estimated credible region spans the Northeast, South, and West regions, excluding the northern and eastern parts of Montana and small portions of Wyoming and Colorado. When considering the joint exceedance level $ u=0.65$, the proportion of the coverage area decreases significantly compared to $u=0.35$. The estimated credible region, choosing $u=0.65$, covers a narrow region in the Northeastern part of the mainland United States, including New Jersey, New York, Pennsylvania, and the District of Columbia, as well as Maryland and some portions of Virginia and North Carolina in the southern region. The estimated hotspot regions include $66\%$ and $9.6\%$ of the total number of grid cells under the two considered thresholds. Although in most regions of the mainland United States the effect of anthropogenic forcing on high-temperature extremes is positive, identifying regions that need immediate action to achieve a sustainable climate may be of high importance to stakeholders and climate policymakers. In this context, to identify the regions that need immediate attention in implementing the regulation policy governing human intervention, hotspot regions are constructed using high thresholds.

\section{Discussion}\label{sec:discussion}

Considering the high applicability of extreme event attribution studies during this alarming phase of climate change, we propose a novel causal metric to quantify the impact of anthropogenic forcing on high-temperature extremes, which will be useful for climatologists working in detection and attribution. This causal metric is developed on top of a unified potential outcome framework that integrates the marginal return levels of annual temperature maxima of both the counterfactual and factual worlds. Besides, it has a valid causal interpretation under unconfounding and in the absence of interference, and we use it to estimate the effect of anthropogenic forcing on high-temperature extremes across the mainland United States.

We employ a latent Gaussian model (LGM), a special class of Bayesian hierarchical model, to correctly specify the marginal return levels and the uncertainties associated with the latent variables and hyperparameters. In the data layer, we model annual temperature maxima using GEV distributions with spatially-varying coefficients, with the temporal component incorporated into the location parameter. We propose a novel transformation for the shape parameter $\xi$ based on three reasonable conditions in the context of temperature extremes. We assert that the transformation method is reliable when specifically applied to high-temperature extremes. From an application perspective and for reliable tail inference, we propose a transformed asymmetric density for $\xi$ \citep{martins2000generalized}. In the latent layer, we employ the ICAR model to capture the conditional neighborhood effect and to smooth the spatial surfaces of posterior estimates. In addition, the conditional autoregressive model features a precision matrix that encodes conditional dependence information, allowing faster Bayesian computation and opening the door to the applicability of sampling-based inferential schemes, such as MCMC methods. We employ an extended version \citep{johannesson2022approximate} of the Max-and-Smooth approach \citep{hrafnkelsson2021max} to bypass the complex posterior computations. The proposed LGM, along with the approximate Bayesian inference scheme, leverages Gibbs sampling to draw posterior inference about the latent variables before other complex MCMC methods. After drawing posterior samples of the corresponding latent parameters used in the causal metric, we obtain the estimated spatial surface of the causal effect over the contiguous United States. A spatial plot of the posterior estimate of the causal effect shows that anthropogenic forcing is more active than other factors in driving frequent high-temperature extremes across the Northeastern region, including most metropolitan cities across the mainland United States. Furthermore, this attribution study unveils the trend in temperature extremes and the effectiveness of anthropogenic forcing in driving it during the pre-industrial period over the mainland United States. Using posterior causal effect estimates and a suitable threshold, the hotspots are constructed over the mainland United States using the approach in \cite{french2016credible}, controlling the family-wise error rate (FWER). The outer credible region corresponding to a high threshold displayed in the study includes the majority of the Northwestern region and a fraction of the Southwestern region, which needs attention. While we focus on the outer credible region, an estimated inner credible region turns out to be a null set; the availability of a dataset with a higher resolution than $1^\circ \times 1^\circ$ would facilitate such an analysis as well.

In our methodology and computation for obtaining efficient marginal return level estimates, implementing approximate rather than exact Bayesian inference and ignoring extremal dependence in the data layer are possible drawbacks. Given that we have a sufficiently large number of temporal observations (165, here) at each grid cell to estimate only four parameters of the marginal distributions, a Laplace approximation of the data likelihood seems reasonable. However, in applications with a very limited number of temporal replications, a Laplace approximation may not be particularly suitable, and replacing the block maxima approach with a peaks-over-threshold or point process approach \citep{hazra2023bayesian} would be a more reasonable option. As noted by \cite{davison2012statistical}, LGMs perform reasonably well for estimating return levels, and thus our methodology may not be suitable for exploring spatial dependence properties. In such a case, a proper max-stable model \citep{schlather2002models, padoan2010likelihood, reich2012hierarchical,huser2024vecchia} would be a more reasonable approach. Another interesting research direction would be incorporating the Whittle-Matérn Brown-Resnick process, recently proposed by \cite{bolin2025intrinsic} for modeling temperature maxima, which is based on a new flexible class of intrinsic Whittle-Mat\'ern Gaussian random fields and facilitates sparse computation. Recently, \cite{cotsakis2026assessing} provided informative metrics based on excursion sets, and metrics inferred from estimated hotspots can provide meaningful insights to policymakers.

\section*{Data and code availability}
The \texttt{R} code and data required to reproduce the figures and tables presented in this article are accessible at the following GitHub repository: \url{https://github.com/ritik-76/Causal_attribution}.

\bibliographystyle{plainnat} 
\bibliography{main}          

@article{kiriliouk2020climate,
  title   = {Climate extreme event attribution using multivariate peaks-over-thresholds modeling and counterfactual theory},
  author  = {Kiriliouk, Anna and Naveau, Philippe},
  journal = {The Annals of Applied Statistics},
  volume  = {14},
  number  = {3},
  pages   = {1342--1358},
  year    = {2020},
  doi     = {10.1214/20-AOAS1355},
  publisher = {Institute of Mathematical Statistics}
}

@article{lee2023ipcc,
  title={{IPCC, 2023: Climate change 2023: Synthesis report, summary for policymakers. Contribution of working groups I, II, and III to the sixth assessment report of the Intergovernmental Panel on Climate Change [core writing team, H. Lee and J. Romero (eds.)]. IPCC, Geneva, Switzerland.}},
  author={Lee, Hoesung and Calvin, Katherine and Dasgupta, Dipak and Krinner, Gerhard and Mukherji, Aditi and Thorne, Peter and Trisos, Christopher and Romero, Jos{\'e} and Aldunce, Paulina and Barret, Ko and others},
  year={2023},
  publisher={Intergovernmental Panel on Climate Change (IPCC)}
}

@article{hazra2020multivariate,
  title={A multivariate spatial skew-t process for joint modeling of extreme precipitation indexes},
  author={Hazra, Arnab and Reich, Brian J and Staicu, Ana-Maria},
  journal={Environmetrics},
  volume={31},
  number={3},
  pages={e2602},
  year={2020},
  publisher={Wiley Online Library}
}

@article{hazra2025efficient,
  title={Efficient modeling of spatial extremes over large geographical domains},
  author={Hazra, Arnab and Huser, Rapha{\"e}l and Bolin, David},
  journal={Journal of Computational and Graphical Statistics},
  volume={34},
  number={3},
  pages={795--811},
  year={2025},
  publisher={Taylor \& Francis}
}

@article{fischer2015anthropogenic,
  title={Anthropogenic contribution to global occurrence of heavy-precipitation and high-temperature extremes},
  author={Fischer, Erich M and Knutti, Reto},
  journal={Nature climate change},
  volume={5},
  number={6},
  pages={560--564},
  year={2015},
  publisher={Nature Publishing Group UK London}
}

@article{zhang2013attributing,
  title={Attributing intensification of precipitation extremes to human influence},
  author={Zhang, Xuebin and Wan, Hui and Zwiers, Francis W and Hegerl, Gabriele C and Min, Seung-Ki},
  journal={Geophysical Research Letters},
  volume={40},
  number={19},
  pages={5252--5257},
  year={2013},
  publisher={Wiley Online Library}
}

@article{stott2016attribution,
  title={Attribution of extreme weather and climate-related events},
  author={Stott, Peter A and Christidis, Nikolaos and Otto, Friederike EL and Sun, Ying and Vanderlinden, Jean-Paul and Van Oldenborgh, Geert Jan and Vautard, Robert and von Storch, Hans and Walton, Peter and Yiou, Pascal and others},
  journal={Wiley Interdisciplinary Reviews: Climate Change},
  volume={7},
  number={1},
  pages={23--41},
  year={2016},
  publisher={Wiley Online Library}
}

@article{naveau2020statistical,
  title={Statistical methods for extreme event attribution in climate science},
  author={Naveau, Philippe and Hannart, Alexis and Ribes, Aur{\'e}lien},
  journal={Annual Review of Statistics and Its Application},
  volume={7},
  number={1},
  pages={89--110},
  year={2020},
  publisher={Annual Reviews}
}

@article{hannart2016causal,
  title={Causal counterfactual theory for the attribution of weather and climate-related events},
  author={Hannart, Alexis and Pearl, Judea and Otto, FEL and Naveau, P and Ghil, M},
  journal={Bulletin of the American Meteorological Society},
  volume={97},
  number={1},
  pages={99--110},
  year={2016}
}

@article{katzfuss2017bayesian,
  title={{A Bayesian hierarchical model for climate change detection and attribution}},
  author={Katzfuss, Matthias and Hammerling, Dorit and Smith, Richard L},
  journal={Geophysical Research Letters},
  volume={44},
  number={11},
  pages={5720--5728},
  year={2017},
  publisher={Wiley Online Library}
}

@article{sang2009hierarchical,
  title={Hierarchical modeling for extreme values observed over space and time},
  author={Sang, Huiyan and Gelfand, Alan E},
  journal={Environmental and Ecological Statistics},
  volume={16},
  number={3},
  pages={407--426},
  year={2009},
  publisher={Springer}
}

@article{cooley2010spatial,
  title={Spatial hierarchical modeling of precipitation extremes from a regional climate model},
  author={Cooley, Daniel and Sain, Stephan R},
  journal={Journal of Agricultural, Biological, and Environmental Statistics},
  volume={15},
  number={3},
  pages={381--402},
  year={2010},
  publisher={Springer}
}

@article{reich2019spatial,
  title={{A spatial Markov model for climate extremes}},
  author={Reich, Brian J and Shaby, Benjamin A},
  journal={Journal of Computational and Graphical Statistics},
  volume={28},
  number={1},
  pages={117--126},
  year={2019},
  publisher={Taylor \& Francis}
}

@article{johannesson2022approximate,
  title={{Approximate Bayesian inference for analysis of spatiotemporal flood frequency data}},
  author={J{\'o}hannesson, {\'A}rni V and Siegert, Stefan and Huser, Rapha{\"e}l and Bakka, Haakon and Hrafnkelsson, Birgir},
  journal={The Annals of Applied Statistics},
  volume={16},
  number={2},
  pages={905--935},
  year={2022},
  publisher={Institute of Mathematical Statistics}
}

@incollection{hazra2023bayesian,
  title={{Bayesian latent Gaussian models for high-dimensional spatial extremes}},
  author={Hazra, Arnab and Huser, Rapha{\"e}l and J{\'o}hannesson, {\'A}rni V},
  booktitle={Statistical Modeling Using Bayesian Latent Gaussian Models: With Applications in Geophysics and Environmental Sciences},
  pages={219--251},
  year={2023},
  publisher={Springer}
}

@article{rubin1978bayesian,
  title={Bayesian inference for causal effects: The role of randomization},
  author={Rubin, Donald B},
  journal={The Annals of Statistics},
  pages={34--58},
  volume={6},
  number={1},
  year={1978},
  publisher={JSTOR}
}

@article{reich2021review,
  title={A review of spatial causal inference methods for environmental and epidemiological applications},
  author={Reich, Brian J and Yang, Shu and Guan, Yawen and Giffin, Andrew B and Miller, Matthew J and Rappold, Ana},
  journal={International Statistical Review},
  volume={89},
  number={3},
  pages={605--634},
  year={2021},
  publisher={Wiley Online Library}
}

@article{davison2015statistics,
  title={Statistics of Extremes},
  author={Davison, Anthony C and Huser, Rapha{\"e}l},
  journal={Annual Review of Statistics and Its Application},
  volume={2},
  number={1},
  pages={203--235},
  year={2015},
  publisher={Annual Reviews Inc.}
}

@book{coles2001introduction,
  title={An introduction to statistical modeling of extreme values},
  author={Coles, Stuart and Bawa, Joanna and Trenner, Lesley and Dorazio, Pat},
  volume={208},
  year={2001},
  publisher={Springer}
}

@article{tawn1988bivariate,
  title={Bivariate extreme value theory: models and estimation},
  author={Tawn, Jonathan A},
  journal={Biometrika},
  volume={75},
  number={3},
  pages={397--415},
  year={1988},
  publisher={Oxford University Press}
}

@article{besag1974spatial,
  title={Spatial interaction and the statistical analysis of lattice systems},
  author={Besag, Julian},
  journal={Journal of the Royal Statistical Society: Series B (Methodological)},
  volume={36},
  number={2},
  pages={192--225},
  year={1974},
  publisher={Wiley Online Library}
}

@article{besag1995conditional,
  title={On conditional and intrinsic autoregressions},
  author={Besag, Julian and Kooperberg, Charles},
  journal={Biometrika},
  volume={82},
  number={4},
  pages={733--746},
  year={1995},
  publisher={Oxford University Press}
}

@article{lavine2012rigorous,
  title={{On rigorous specification of ICAR models}},
  author={Lavine, Michael L and Hodges, James S},
  journal={The American Statistician},
  volume={66},
  number={1},
  pages={42--49},
  year={2012},
  publisher={Taylor \& Francis}
}

@article{gelfand2003proper,
  title={Proper multivariate conditional autoregressive models for spatial data analysis},
  author={Gelfand, Alan E and Vounatsou, Penelope},
  journal={Biostatistics},
  volume={4},
  number={1},
  pages={11--15},
  year={2003},
  publisher={Oxford University Press}
}

@article{larsen2022spatial,
  title={{A spatial causal analysis of wildland fire-contributed PM2. 5 using numerical model output}},
  author={Larsen, Alexandra and Yang, Shu and Reich, Brian J and Rappold, Ana G},
  journal={The Annals of Applied Statistics},
  volume={16},
  number={4},
  pages={2714},
  year={2022}
}

@article{patil2010digital,
  title={Digital governance, hotspot geoinformatics, and sustainable development: A Preface},
  author={Patil, Ganapati P},
  journal={Environmental and Ecological Statistics},
  volume={17},
  number={2},
  pages={133},
  year={2010},
  publisher={Springer Nature BV}
}

@article{french2013spatio,
  title={Spatio-temporal exceedance locations and confidence regions},
  author={French, Joshua P and Sain, Stephan R},
  journal={The Annals of Applied Statistics},
  year={2013}
}

@article{furrer2007spatial,
  title={{Spatial patterns of probabilistic temperature change projections from a multivariate Bayesian analysis}},
  author={Furrer, Reinhard and Knutti, Reto and Sain, SR and Nychka, DW and Meehl, GA},
  journal={Geophysical Research Letters},
  volume={34},
  number={6},
  year={2007},
  publisher={Wiley Online Library}
}

@article{sain2011spatial,
  title={A spatial analysis of multivariate output from regional climate models},
  author={Sain, Stephan R and Furrer, Reinhard and Cressie, Noel},
  journal={The Annals of Applied Statistics},
  pages={150--175},
  volume={5},
  number={1},
  year={2011},
  publisher={JSTOR}
}

@article{bolin2009fast,
  title={{Fast estimation of spatially dependent temporal vegetation trends using Gaussian Markov random fields}},
  author={Bolin, David and Lindstr{\"o}m, Johan and Eklundh, Lars and Lindgren, Finn},
  journal={Computational Statistics \& Data Analysis},
  volume={53},
  number={8},
  pages={2885--2896},
  year={2009},
  publisher={Elsevier}
}

@article{eklundh2003vegetation,
  title={{Vegetation index trends for the African Sahel 1982--1999}},
  author={Eklundh, Lars and Olsson, Lennart},
  journal={Geophysical Research Letters},
  volume={30},
  number={8},
  year={2003},
  publisher={Wiley Online Library}
}

@inproceedings{cressie2005geostatistical,
  title={Geostatistical prediction of spatial extremes and their extent},
  author={Cressie, N and Zhang, J and Craigmile, PF},
  booktitle={Geostatistics for Environmental Applications: Proceedings of the Fifth European Conference on Geostatistics for Environmental Applications},
  pages={27--37},
  year={2005},
  organization={Springer}
}

@article{bolin2015excursion,
  title={{Excursion and contour uncertainty regions for latent Gaussian models}},
  author={Bolin, David and Lindgren, Finn},
  journal={Journal of the Royal Statistical Society Series B: Statistical Methodology},
  volume={77},
  number={1},
  pages={85--106},
  year={2015},
  publisher={Oxford University Press}
}

@article{french2016credible,
  title={Credible regions for exceedance sets of geostatistical data},
  author={French, Joshua P and Hoeting, Jennifer A},
  journal={Environmetrics},
  volume={27},
  number={1},
  pages={4--14},
  year={2016},
  publisher={Wiley Online Library}
}

@article{hazra2021estimating,
  title={{Estimating high-resolution Red Sea surface temperature hotspots, using a low-rank semiparametric spatial model}},
  author={Hazra, Arnab and Huser, Rapha{\"e}l},
  journal={The Annals of Applied Statistics},
  volume={15},
  number={2},
  pages={572--596},
  year={2021},
  publisher={Institute of Mathematical Statistics}
}

@article{gilbert2021causal,
  title={A causal inference framework for spatial confounding},
  author={Gilbert, Brian and Datta, Abhirup and Casey, Joan A and Ogburn, Elizabeth L},
  journal={arXiv preprint arXiv:2112.14946},
  year={2021}
}

@article{papadogeorgou2023spatial,
  title={Spatial causal inference in the presence of unmeasured confounding and interference},
  author={Papadogeorgou, Georgia and Samanta, Srijata},
  journal={arXiv preprint arXiv:2303.08218},
  year={2023}
}

@article{boucher2020presentation,
  title={{Presentation and evaluation of the IPSL-CM6A-LR climate model}},
  author={Boucher, Olivier and Servonnat, J{\'e}r{\^o}me and Albright, Anna Lea and Aumont, Olivier and Balkanski, Yves and Bastrikov, Vladislav and Bekki, Slimane and Bonnet, R{\'e}my and Bony, Sandrine and Bopp, Laurent and others},
  journal={Journal of Advances in Modeling Earth Systems},
  volume={12},
  number={7},
  pages={e2019MS002010},
  year={2020},
  publisher={Wiley Online Library}
}

@article{giffin2023generalized,
  title={Generalized propensity score approach to causal inference with spatial interference},
  author={Giffin, Andrew and Reich, BJ and Yang, Shu and Rappold, AG},
  journal={Biometrics},
  volume={79},
  number={3},
  pages={2220--2231},
  year={2023},
  publisher={Oxford University Press}
}

@article{bhowmik2025bayesian,
  title={{A Bayesian latent Gaussian conditional autoregressive copula model for analyzing spatially-varying trends in rainfall}},
  author={Bhowmik, Sayan and Hazra, Arnab},
  journal={Japanese Journal of Statistics and Data Science},
  year={2026},
  pages={99--131},
  volume={9},
  number={1}
}

@book{cressie2011statistics,
  title={Statistics for spatio-temporal data},
  author={Cressie, Noel and Wikle, Christopher K},
  year={2011},
  publisher={John Wiley \& Sons}
}

@book{banerjee2003hierarchical,
  title={Hierarchical modeling and analysis for spatial data},
  author={Banerjee, Sudipto and Carlin, Bradley P and Gelfand, Alan E},
  year={2003},
  publisher={Chapman and Hall/CRC}
}

@inproceedings{fisher1928limiting,
  title={Limiting forms of the frequency distribution of the largest or smallest member of a sample},
  author={Fisher, Ronald Aylmer and Tippett, Leonard Henry Caleb},
  booktitle={Mathematical proceedings of the Cambridge philosophical society},
  volume={24},
  number={2},
  pages={180--190},
  year={1928},
  organization={Cambridge University Press}
}

@article{husler1989maxima,
  title={Maxima of normal random vectors: between independence and complete dependence},
  author={H{\"u}sler, J{\"u}rg and Reiss, Rolf-Dieter},
  journal={Statistics \& Probability Letters},
  volume={7},
  number={4},
  pages={283--286},
  year={1989},
  publisher={Elsevier}
}

@article{sibuya1960bivariate,
  title={Bivariate extreme statistics},
  author={Sibuya, Masaaki and others},
  journal={Annals of the Institute of Statistical Mathematics},
  volume={11},
  number={2},
  pages={195--210},
  year={1960},
  publisher={Tokyo}
}

@article{stephenson2018evd,
  title={evd: Functions for extreme value distributions},
  author={Stephenson, Alec and Ferro, Chris},
  journal={R package version},
  volume={2},
  pages={3--3},
  year={2018}
}

@inproceedings{hasan2013modeling,
  title={{Modeling annual extreme temperature using generalized extreme value distribution: A case study in Malaysia}},
  author={Hasan, Husna and Salam, Norfatin and Kassim, Suraiya},
  booktitle={AIP Conference Proceedings},
  volume={1522},
  number={1},
  pages={1195--1203},
  year={2013},
  organization={American Institute of Physics}
}

@article{hogan2019representation,
  title={{Representation of US warm temperature extremes in global climate model ensembles}},
  author={Hogan, Emily and Nicholas, Robert E and Keller, Klaus and Eilts, Stephanie and Sriver, Ryan L},
  journal={Journal of Climate},
  volume={32},
  number={9},
  pages={2591--2603},
  year={2019}
}

@article{martins2000generalized,
  title={Generalized maximum-likelihood generalized extreme-value quantile estimators for hydrologic data},
  author={Martins, Eduardo S and Stedinger, Jery R},
  journal={Water Resources Research},
  volume={36},
  number={3},
  pages={737--744},
  year={2000},
  publisher={Wiley Online Library}
}

@article{neyman1923applications,
  title={Sur les applications de la th{\'e}orie des probabilit{\'e}s aux experiences agricoles: Essai des principes},
  author={Neyman, Jersey},
  journal={Roczniki Nauk Rolniczych},
  volume={10},
  number={1},
  pages={1--51},
  year={1923}
}

@article{holland1986statistics,
  title={Statistics and causal inference},
  author={Holland, Paul W},
  journal={Journal of the American statistical Association},
  volume={81},
  number={396},
  pages={945--960},
  year={1986},
  publisher={Taylor \& Francis}
}

@article{hrafnkelsson2012spatial,
  title={{Spatial modeling of annual minimum and maximum temperatures in Iceland}},
  author={Hrafnkelsson, Birgir and Morris, Jeffrey S and Baladandayuthapani, Veerabhadran},
  journal={Meteorology and Atmospheric Physics},
  volume={116},
  number={1},
  pages={43--61},
  year={2012},
  publisher={Springer}
}

@book{reich2019bayesian,
  title={Bayesian statistical methods},
  author={Reich, Brian J and Ghosh, Sujit K},
  year={2019},
  publisher={Chapman and Hall/CRC}
}

@article{hrafnkelsson2021max,
  title={{Max-and-smooth: A two-step approach for approximate Bayesian inference in latent Gaussian models}},
  author={Hrafnkelsson, Birgir and Siegert, Stefan and Huser, Rapha{\"e}l and Bakka, Haakon and J{\'o}hannesson, {\'A}rni V},
  journal={Bayesian Analysis},
  volume={16},
  number={2},
  pages={611--638},
  year={2021},
  publisher={International Society for Bayesian Analysis}
}

@article{furrer2010spam,
  title={{spam: A sparse matrix R package with emphasis on MCMC methods for Gaussian Markov random fields}},
  author={Furrer, Reinhard and Sain, Stephan R},
  journal={Journal of Statistical Software},
  volume={36},
  pages={1--25},
  year={2010}
}

@book{hrafnkelsson2023statistical,
  title={Statistical Modeling Using Bayesian Latent Gaussian Models: With Applications in Geophysics and Environmental Sciences},
  author={Hrafnkelsson, Birgir},
  year={2023},
  publisher={Springer}
}

@article{huang2016estimating,
  title={{Estimating changes in temperature extremes from millennial-scale climate simulations using generalized extreme value (GEV) distributions}},
  author={Huang, Whitney K and Stein, Michael L and McInerney, David J and Sun, Shanshan and Moyer, Elisabeth J},
  journal={Advances in Statistical Climatology, Meteorology and Oceanography},
  volume={2},
  number={1},
  pages={79--103},
  year={2016},
  publisher={Copernicus Publications G{\"o}ttingen, Germany}
}

@article{cooley2009extreme,
  title={Extreme value analysis and the study of climate change: A commentary on Wigley 1988},
  author={Cooley, Daniel},
  journal={Climatic Change},
  volume={97},
  number={1},
  pages={77--83},
  year={2009},
  publisher={Springer}
}

@article{andresen2012historical,
  title={{Historical climate and climate trends in the Midwestern USA}},
  author={Andresen, Jeff and Hilberg, Steve and Kunkel, Ken and Center, Midwest Regional Climate},
  journal={US National Climate Assessment Midwest Technical Input Report},
  pages={1--18},
  year={2012}
}

@article{fonseca2000changing,
  title={{Changing patterns of population density in the United States}},
  author={Fonseca, James W and Wong, David W},
  journal={The Professional Geographer},
  volume={52},
  number={3},
  pages={504--517},
  year={2000},
  publisher={Wiley Online Library}
}

@article{pryor2013midwestern,
  title={{The Midwestern United States: Socio-economic context and physical climate}},
  author={Pryor, Sara C and Barthelmie, Rebecca J},
  journal={Climate Change in the Midwest: Impacts, Risks, Vulnerability, and Adaptation, Indiana University Press, Bloomington, IN},
  pages={12--47},
  year={2013}
}

@article{hartley2021preliminary,
  title={{A preliminary analysis of US and state-level results from the 2020 census}},
  author={Hartley, Christine and Perry, Marc and Rogers, Luke},
  journal={US Census Bureau},
  year={2021}
}

@article{davison2012statistical,
  title={Statistical modeling of spatial extremes},
  author={Davison, Anthony C. and Padoan, Simone A. and Ribatet, Mathieu},
  journal={Statistical Science},
  volume={27},
  number={2},
  pages={161--186},
  year={2012},
  publisher={Institute of Mathematical Statistics},
  doi={10.1214/11-STS376}
}

@book{cressie2015statistics,
  title={Statistics for spatial data},
  author={Cressie, Noel},
  year={2015},
  publisher={John Wiley \& Sons}
}

@article{bolin2025intrinsic,
  title={{Intrinsic Whittle--Mat\'ern fields and sparse spatial extremes}},
  author={Bolin, David and Braunsteins, Peter and Engelke, Sebastian and Huser, Rapha{\"e}l},
  journal={arXiv preprint arXiv:2512.23395},
  year={2025}
}

@article{cotsakis2026assessing,
  title={Assessing the size of spatial extreme events using local coefficients based on excursion sets},
  author={Cotsakis, Ryan and Di Bernardino, Elena and Opitz, Thomas},
  journal={Spatial Statistics},
  pages={100958},
  year={2026},
  volume = {72},
  publisher={Elsevier}
}

@article{schlather2002models,
  title={Models for stationary max-stable random fields},
  author={Schlather, Martin},
  journal={Extremes},
  volume={5},
  number={1},
  pages={33--44},
  year={2002},
  publisher={Springer}
}

@article{reich2012hierarchical,
  title={A hierarchical max-stable spatial model for extreme precipitation},
  author={Reich, Brian J and Shaby, Benjamin A},
  journal={The Annals of Applied Statistics},
  volume={6},
  number={4},
  pages={1430},
  year={2012}
}

@article{huser2024vecchia,
  title={Vecchia likelihood approximation for accurate and fast inference with intractable spatial max-stable models},
  author={Huser, Rapha{\"e}l and Stein, Michael L and Zhong, Peng},
  journal={Journal of Computational and Graphical Statistics},
  volume={33},
  number={3},
  pages={978--990},
  year={2024},
  publisher={Taylor \& Francis}
}

@article{padoan2010likelihood,
  title={Likelihood-based inference for max-stable processes},
  author={Padoan, Simone A and Ribatet, Mathieu and Sisson, Scott A},
  journal={Journal of the American Statistical Association},
  volume={105},
  number={489},
  pages={263--277},
  year={2010},
  publisher={Taylor \& Francis}
}

\end{document}